\documentclass[aps,twocolumn,pra,superscript,floatfix,superscriptaddress,showpacs]{revtex4-1}
\usepackage{amssymb}

\usepackage{epsf}
\usepackage{epsfig}
\usepackage{graphicx}
\usepackage{dcolumn}
\usepackage{braket}
\usepackage{bm}
\usepackage{amsfonts}
\usepackage{amsmath}
\usepackage{amssymb}
\usepackage{color,soul}
\usepackage{wasysym}
\usepackage{mathrsfs}
\usepackage{natbib}
\usepackage{float}
\usepackage{multirow}
\usepackage[colorlinks=true,%
            linkcolor=blue,%
            urlcolor=blue,%
            citecolor=blue,%
            filecolor=blue,%
            bookmarksopen=true,%
            pdfauthor={J. P. Ramos-Andrade},%
            pdftitle={ProICPSarxiv2},%
            pdfsubject={ProICPSarxiv2},%
            pdfpagemode=UseOutlines]{hyperref}

\newcommand{\bea}{\begin{eqnarray}}
\newcommand{\eea}{\end{eqnarray}}

\begin{document}

\title{Fano-Majorana effect and bound states in the continuum on a crossbar-shaped quantum dot hybrid structure}

\author{J. P. Ramos-Andrade}
\email{juan.ramosa@usm.cl}	
\affiliation{Departamento de F\'isica, Universidad T\'ecnica Federico Santa Mar\'ia, Casilla 110 V, Valpara\'iso, Chile}
\author{D. Zambrano}
\affiliation{Departamento de F\'isica, Universidad T\'ecnica Federico Santa Mar\'ia, Casilla 110 V, Valpara\'iso, Chile}
\author{P. A. Orellana}
\affiliation{Departamento de F\'isica, Universidad T\'ecnica Federico Santa Mar\'ia, Casilla 110 V, Valpara\'iso, Chile}

\begin{abstract}
We investigate transport properties through a crossbar-shaped structure formed by a quantum dot (QD) coupled to two normal leads and embedded between two one-dimensional topological superconductors (TSCs). Each TSC hosts Majorana bound states (MBSs) at its ends, which can interact between them with an effective coupling strength. We find a signature of bound states in continuum (BIC) in the MBSs spectral function. By allowing finite inter MBSs coupling, BICs splitting is observed and shows projection in transmission for asymmetric coupling case as cuasi-BICs. As a consequence, we also show that the Fano effect, arising from interference phenomena between MBSs hybridization trough QD, is observed with a half-integer amplitude modulation. We believe our findings can help to better understand the properties of MBSs and their interplay with QDs.
\end{abstract}

\maketitle

\section{Introduction}
\label{introduction}

The possible realization of exotic quasiparticles like anyons in solid state systems, as bound states with zero energy satisfying non-Abelian statistics, have been attracting attention due to their promising applications in quantum computing. One of them was first predicted by E. Majorana \cite{Majorana1937}, which has as a principal feature to be its own antiparticle. In the last decade, Majorana fermions (MFs) have become a hot topic in condensed matter physics \cite{Wilczek2009,Franz2010,Wu2012} and quantum computation \cite{Nayak2008,Leijnse2011,Bravyi2000,Kitaev2001,Kitaev2003,Pachos2012,Kraus2013,Albrecht2016}, since they can be manipulated with braiding operations \cite{Kraus2013}, allowing to perform fault-tolerant quantum gates \cite{Nayak2008,Kitaev2001,Pachos2012,Albrecht2016,Beenakker2013,Laflamme2014}. A qubit built with this exotic quasiparticles is topologically protected when localized MFs, Majorana bound states (MBSs), are spatially separated. i.e., unpaired. Among other systems, MBSs are predicted to be found at the ends of a topological one-dimensional semiconductor-superconductor nanowire with strong spin-orbit interaction in the presence of a magnetic field, namely topological superconductor (TSC) \cite{Moore2009}. This system can be seen as an implementation of a Kitaev chain \cite{Kitaev2001}, in which the coupling between the two MBSs is expected to decay exponentially with wire length \cite{Albrecht2016}, protecting the qubit from decoherence by local perturbations \cite{Wu2012,Kitaev2001,Kraus2013,Albrecht2016,Semenoff2006,Tewari2008}.

One of the main challenges is pointing out to detection of the existence of MBSs, as well as their characterization. Many systems have been proposed through the literature \cite{Law2009,Pikulin2012,Franz2013,Prada2012,Rainis2013,Cook2012,Liu2013,Stanescu2011,Lee2014,Flensberg2010,Wimmer2011,Nilsson2008,Bolech2007,Fu2009}, and several experiments have been carried out based on zero-bias anomalies in transport properties through source/drain leads \cite{Mourik2012,Deng2012,Das2012,Lee2012,Finck2013,Churchill2013}, but not all of these anomalies are evidence of MBS. For instance, at zero energy and low-temperature other phenomenology could take place, as Kondo effect \cite{Goldhaber1998,Cronenwett1998} and Andreev bound states, where the latter is due to the electron and hole scattering at the normal-superconductor interface \cite{Golubov2009}. Quantum dots (QDs) have shown to have rich interference phenomena to exploit when multiple QDs structures are considered. These structures \cite{Holleitner2001,Holleitner2002,Shangguan2001,Orellana2003} show the two most important aspects that make them a useful candidate to build nanodevices. In the first place, the possibility to tune a large number of parameters present the system, and the second one, a rich quantum interference mechanisms due to the interaction between the different discretized QDs energy levels, e.g. the Fano effect \cite{Fano1961,Miroshnichenko2010}. These two main aspects will give rise to far more complex quantum transport patterns in hybrid multiple QDs-TSCs structures, making them good candidates to establish MBSs properties in the system \cite{Vernek2014,Ruiz2015,Gong2014,Orellana2003,Meir1992}. In non-interacting QD-leads systems with a side coupled QD, an special signature of the presence of MBSs was established by Liu and Baranger, which is a half-integer conductance at zero energy \cite{Liu2011}. Later, Vernek \textit{et al}. \cite{Vernek2014} have shown that this zero-bias anomaly is due to MBS leaking into the QD and it is robustly pinned against changes in QD energy level, being recently verified \cite{Deng2016}.

On the other hand, new properties are present in quantum interference systems, for instance, some states do not decay even if their energy levels are within the range of the continuum states \cite{bics}, the so-called bound states in the continuum (BICs). The BICs were predicted by von Neumann and Wigner in the dawn of the quantum mechanics \cite{vonNeumann1929}. Recently, the interest in the investigation regarding BICs due to the observation of this kind of states in photonic systems. Since interference phenomena take place in electronic systems in analogy with the photonic ones, the inherent possibility of the presence of BICs emerges \cite{Hsu2016,Ramos2014}. In QDs-MBSs systems, a theoretical encryption device based on BICs \cite{Guessi2017} and Majorana qubit readout technology \cite{Ricco2016} has been proposed.

In a previous work, we proposed a combined system of multiple-QDs and MBSs, which is capable of veil/unveil BICs due to the interaction with MBSs \cite{Zambrano2018}. Using tunable gate voltages \cite{Alicea2011} the topological properties of the MBSs can be manipulated allowing them to protect the information stored in the BIC. In the present work, we study a system form by a single-QD embedded between current leads and connected to TSCs hosting MBSs at its ends, finding interesting features in this simplified system. We focus on QD density of states and MBSs spectral function calculated through Green's functions to identify signatures in transmission probability, with respect to the QD-MBSs coupling. Our results show that the energy localization of BICs and their widths can be controlled by tuning the inter-MBSs coupling in each TSC for the case with QD energy level aligned with Fermi energy. By setting the QD energy level above/below Fermi energy, the leaking of the BICs into the transmission leads to an amplitude modified Fano effect. We believe our findings could be useful to give a further characterization of MBSs in interplay with a QD.

This paper is organized by presenting the model and the corresponding Hamiltonian with the method considered to obtain quantities of interest in Section \ref{Model}; Section \ref{Results} shows the results and the corresponding discussion, and finally, the concluding remarks are presented in Section \ref{Summary}.

\section{Model}
\label{Model}

\begin{figure}[!h]
\centering
\includegraphics[width=1.0\linewidth]{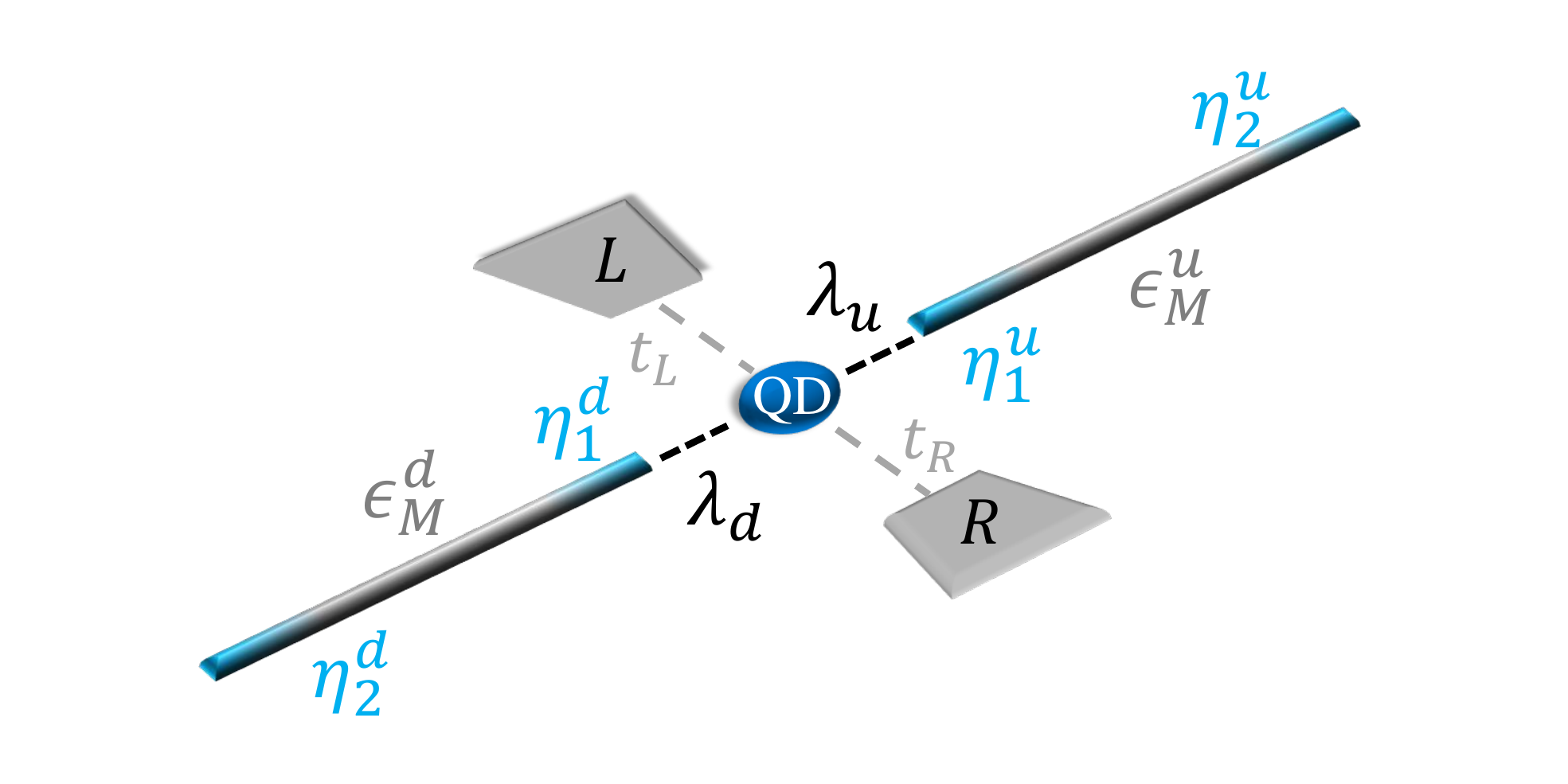}
\caption{Model setup: Crossbar-shaped TSC-QD-TSC system. QD (blue) coupled to two normal leads (solid gray), and two TSCs (gray tones) $u$ and $d$, each hosting two MBSs (light blue), $\eta_{_1}^{u(d)}$ and $\eta_{_2}^{u(d)}$.}
\label{fig.montaje}
\end{figure}

We consider a crossbar-shaped structure form by a QD, two normal leads and two TSCs hosting MBSs at its ends. The system is such as the QD is connected with both leads, labeled as $L$ and $R$, and with both side-coupled TSCs, as we show schematically in Fig.\ \ref{fig.montaje}. We model the system with an effective low-energy Hamiltonian in the following form,
\begin{equation}
H = H_{\text{leads}} + H_{\text{dot}} + H_\text{dot-leads} + H_{\text{dot-M}} + H_\text{M}\,,\label{H}
\end{equation}			
where the first three terms on the right-hand side correspond to normal leads, the QD and the connection between them, respectively. These are given by
\begin{equation}
H_{\text{leads}} = \sum_{\alpha,{\bf k}}{ \varepsilon_{\alpha, {\bf k}} c_{\alpha, {\bf k}}^\dagger c_{\alpha, {\bf k}} }\,\text{,}
\end{equation}
\begin{equation}
H_\text{dot} = \varepsilon_{d} d^\dagger d \text{,}
\end{equation}
\begin{equation}
H_\text{dot-leads}  =  \sum_{\alpha,{\bf k}}{t_{\alpha} d^\dagger c_{\alpha, {\bf k}} } + \text{h.c.} \,\text{,}
\end{equation}
where $c_{\alpha, {\bf k}}^\dagger ( c_{\alpha, {\bf k}} )$ is the electron creation (annihilation) operator with momentum {\bf k} and energy $\varepsilon_{\alpha, {\bf k}}$ in the lead $\alpha = L,R$. $d^\dagger (d)$ is electron creation (annihilation) operator in the QD, with single energy level $\varepsilon_{d}$. $t_{\alpha}$ is the {\bf k}-independent dot-lead tunneling coupling.
			
The last two terms in Eq.\ (\ref{H}) correspond to MBSs and their couplings with the QD, respectively. They are given by
\begin{eqnarray}
H_\text{dot-M} &=&  \left( \lambda_{d}d - \lambda_{d}^{\ast}d^\dagger \right) \eta_{1}^d + \left( \lambda_{u}d - \lambda_{u}^{\ast}d^\dagger \right) \eta_{1}^u \text{,}\label{HdotM}\\
H_\text{M} &=& i\epsilon_{_{M}}^d\eta_{1}^d\eta_{2}^d + i\epsilon_{_{M}}^u\eta_{1}^u\eta_{2}^u\label{HM} \text{,}
\end{eqnarray}
where $\eta_{\beta}^{d(u)}$ denotes the MBS operator, which satisfies both $\eta_{\beta}^{d(u)}=\left[\eta_{\beta}^{d(u)}\right]^{\dag}$ and $\left\{\eta_{\beta}^{d(u)},\eta_{\beta'}^{d(u)}\right\}=\delta_{\beta,\beta'}$ with $\beta=1,2$. Besides, $\lambda_{{d(u)}}$ is the tunneling coupling between $\eta_{1}^{d(u)}$ and the QD, and $\epsilon_{{M}}^{d(u)} \propto \exp(-L_{d(u)}/\zeta)$ is the coupling strength between two MBS in the same TSC, with length $L_{d(u)}$, being $\zeta$ the superconducting coherence length.

A useful way to treat the system analytically is by writing each MBS as a superposition of regular fermionic operators as follows,

\begin{subequations}
\begin{eqnarray}
\eta_{1}^{d(u)}&=&\frac{1}{\sqrt{2}}\left(f_{d(u)}+f_{d(u)}^{\dag}\right)\,,\\
\eta_{2}^{d(u)}&=&-\frac{i}{\sqrt{2}}\left(f_{d(u)}-f_{d(u)}^{\dag}\right)\,,
\end{eqnarray}		
\end{subequations}
which satisfy the following anti-commutation relations $\left\{ f_{\nu},f_{\nu} \right\} = \left\{ f_{\nu}^{\dag},f_{\nu}^{\dag} \right\} = 0$ and $\left\{ f_{\nu},f_{\nu}^{\dag} \right\} = 1$ ($\nu=u,d$). In addition, without loss of generality, we fixed $\lambda_{d}=\lambda_{d}^{\ast}$ and $\lambda_{u}=|\lambda_{u}|\exp(i\theta/2)$, where $\theta$ represents the phase difference between the two TSCs. According to this, Eqs.\ (\ref{HdotM}) and (\ref{HM}) transform to
\begin{eqnarray}
H_\text{dot-M} &=& \frac{\left|\lambda_d\right|}{\sqrt{2}}\left( d - d^\dagger \right) \left(f_{d}+f_{d}^{\dag}\right)\nonumber\\ &+& \frac{\left|\lambda_u\right|}{\sqrt{2}}\left(e^{i\theta/2}d - e^{-i\theta/2}d^\dagger \right) \left(f_{u}+f_{u}^{\dag}\right)\text{,}
\label{HdotMf}\\
H_\text{M} &=& \epsilon_{_{M}}^d f_{d}^{\dag}f_{d} + \epsilon_{_{M}}^u f_{u}^{\dag}f_{u}\label{HMf}\,.
\end{eqnarray}

The leads contribution is included as a self-energy $\Sigma_{\alpha}^{e(h)}(\varepsilon)$ for electrons(holes). In the wide-band limit approximation it is energy-independent, such that $\Sigma_{\alpha}^{e(h)}\equiv-i\Gamma_{\alpha}^{e(h)}$, and fulfills electron hole symmetry, hence $\Gamma_{\alpha}^{e}=\Gamma_{\alpha}^{h}\equiv\Gamma_{\alpha}$. We consider symmetric QD-leads coupling $\Gamma_{\alpha}\equiv\Gamma/2$, so $\Gamma_{{L}}+\Gamma_{{R}}=\Gamma$. In this scenario, the retarded Green's function of the system adopt the matrix form
\begin{align}
&\left[\mathbf{G}^R\right]^{-1} =\nonumber\\
&\left(
\begin{smallmatrix}
  	g_{_M}^u(\varepsilon)^{-1}                                       &  0                                                      &  \frac{\left|\lambda_u\right|}{\sqrt{2}}e^{i\frac{\theta}{2}}  &  -\frac{\left|\lambda_u\right|}{\sqrt{2}}e^{-i\frac{\theta}{2}}  &  0                                          &  0                                         \\
  	0                                                      &  \widetilde{g}_{_M}^u(\varepsilon)^{-1}                           &  \frac{\left|\lambda_u\right|}{\sqrt{2}}e^{i\frac{\theta}{2}}  &  -\frac{\left|\lambda_u\right|}{\sqrt{2}}e^{-i\frac{\theta}{2}}  &  0                                          &  0                                         \\
  	\frac{\left|\lambda_u\right|}{\sqrt{2}}e^{-i\frac{\theta}{2}}  &  \frac{\left|\lambda_u\right|}{\sqrt{2}}e^{-i\frac{\theta}{2}}  &  g_{d}(\varepsilon)^{-1}                                        &  0                                                       &   \frac{\left|\lambda_d\right|}{\sqrt{2}}   &   \frac{\left|\lambda_d\right|}{\sqrt{2}}  \\
  	-\frac{\left|\lambda_u\right|}{\sqrt{2}}e^{i\frac{\theta}{2}}  &  -\frac{\left|\lambda_u\right|}{\sqrt{2}}e^{i\frac{\theta}{2}}  &  0                                                     &  \widetilde{g}_{d}(\varepsilon)^{-1}                              &  -\frac{\left|\lambda_d\right|}{\sqrt{2}}   &  -\frac{\left|\lambda_d\right|}{\sqrt{2}}  \\
  	0                                                      &  0                                                      &  \frac{\left|\lambda_d\right|}{\sqrt{2}}               &  -\frac{\left|\lambda_d\right|}{\sqrt{2}}                &  g_{_M}^d(\varepsilon)^{-1}                           &  0                                         \\
  	0                                                      &  0                                                      &  \frac{\left|\lambda_d\right|}{\sqrt{2}}               &  -\frac{\left|\lambda_d\right|}{\sqrt{2}}                &  0                                          &  \widetilde{g}_{_M}^d(\varepsilon)^{-1}
\end{smallmatrix}
\right)
\text{,}
\label{matrix}
\end{align}
where the diagonal matrix elements are given by
\begin{eqnarray}
	g_{d}(\varepsilon)^{-1}                    &=& \varepsilon -\varepsilon_{d}+i\Gamma\text{,}\\
	\widetilde{g}_{d}(\varepsilon)^{-1}        &=& \varepsilon +\varepsilon_{d}+i\Gamma\text{,}\\
	g_{_M}^{u(d)}(\varepsilon)^{-1}             &=& \varepsilon -\epsilon_{_M}^{u(d)}+i0^+\text{,}\\
	\widetilde{g}_{_M}^{u(d)}(\varepsilon)^{-1} &=& \varepsilon +\epsilon_{_M}^{u(d)}+i0^+\text{,}
\end{eqnarray}
being $0^{+}$ an infinitesimal positive number. The transmission probability for our symmetric leads coupling can be written out as
\begin{equation}
	T(\varepsilon) = -\Gamma\,\text{Im}\left[G_{d}^R(\varepsilon)\right] \text{,}\label{tras}
\end{equation}
where $G_{d}^R$ is the QD retarded Green function. Similarly, the local density of states (LDOS) for the QD can be also expressed in terms of $G_{d}^R$ as
\begin{equation}\label{LDOS}
	\text{LDOS}_{d}(\varepsilon) = -\frac{1}{\pi}\text{Im}\left[G_{d}^R(\varepsilon)\right] \text{,}
\end{equation}
and the spectral function for MBSs is given by
\begin{equation}
	A_{\beta}^{\nu}(\varepsilon) = -2\,\text{Im}\left[G_{{\beta,\nu}}^R(\varepsilon)\right] \text{,}
\end{equation}
where $\nu=u,d$.

The Green's function element for the QD present in Eqs. (\ref{tras}) and (\ref{LDOS}), is obtained analytically using the equation of motion (EOM) procedure. Then, in the energy domain, is given by

\begin{align}
&\left[G_{d}^{R}(\varepsilon)\right]^{-1}=\varepsilon-\varepsilon_{d}+i\Gamma-\lambda^{2}(K_{d}(\varepsilon)+K_{u}(\varepsilon))\\
&-\lambda^{4}\left(\frac{K_{d}^{2}(\varepsilon)+K_{u}^{2}(\varepsilon)+2\cos(\theta)K_{d}(\varepsilon)K_{u}(\varepsilon)}{\varepsilon+\varepsilon_{d}+i\Gamma-\lambda^{2}(K_{d}(\varepsilon)+K_{u}(\varepsilon))}\right)\nonumber\,,
\end{align}
where we have considered symmetric MBS-QD couplings ($|\lambda_{u(d)}| = \lambda$) and
\begin{equation}
K_{\nu}(\varepsilon)=\frac{\varepsilon}{\left(\varepsilon+\epsilon_{_M}^{\nu}\right)\left(\varepsilon-\epsilon_{_M}^{\nu}\right)}\,.
\end{equation}

On the other hand, the full Green's function poles are closely related with the eigenvalues of the isolated Hamiltonian (disconnected from leads) and give reliable information about energy localization of the states. As we consider $|\lambda_{u(d)}| = \lambda$, for $\varepsilon_{d} = 0$, the system eigenvalues are given by

\begin{eqnarray}
  \varepsilon_{0}^{\pm}&=&0\,,\label{E0}\\
  2(\varepsilon_{1}^{\pm})^{2}&=&(\epsilon_{_M}^{u})^{2}+(\epsilon_{_M}^{d})^{2}+4\lambda^{2}\nonumber\\&-&\sqrt{\left[(\epsilon_{_M}^{u})^{2}-(\epsilon_{_M}^{d})^{2}\right]^{2}+8\lambda^{4}(1+\cos(\theta))}\,,\label{E1}\\
  2(\varepsilon_{2}^{\pm})^{2}&=&(\epsilon_{_M}^{u})^{2}+(\epsilon_{_M}^{d})^{2}+4\lambda^{2}\nonumber\\&+&\sqrt{\left[(\epsilon_{_M}^{u})^{2}-(\epsilon_{_M}^{d})^{2}\right]^{2}+8\lambda^{4}(1+\cos(\theta))}\,,\label{E2}
\end{eqnarray}
where $\varepsilon_{0}^{\pm}$ has doubly degeneracy.

\section{Results}\label{Results}

The following results are performed at temperature $\mathcal{T}=0$, and we adopt the energy parameter $\Gamma$ as energy unit throughout the manuscript.

\subsection{Without phase difference, $\theta=0$}\label{theta0}

\begin{figure}[!b]
\centering
\includegraphics[width=1.0\linewidth]{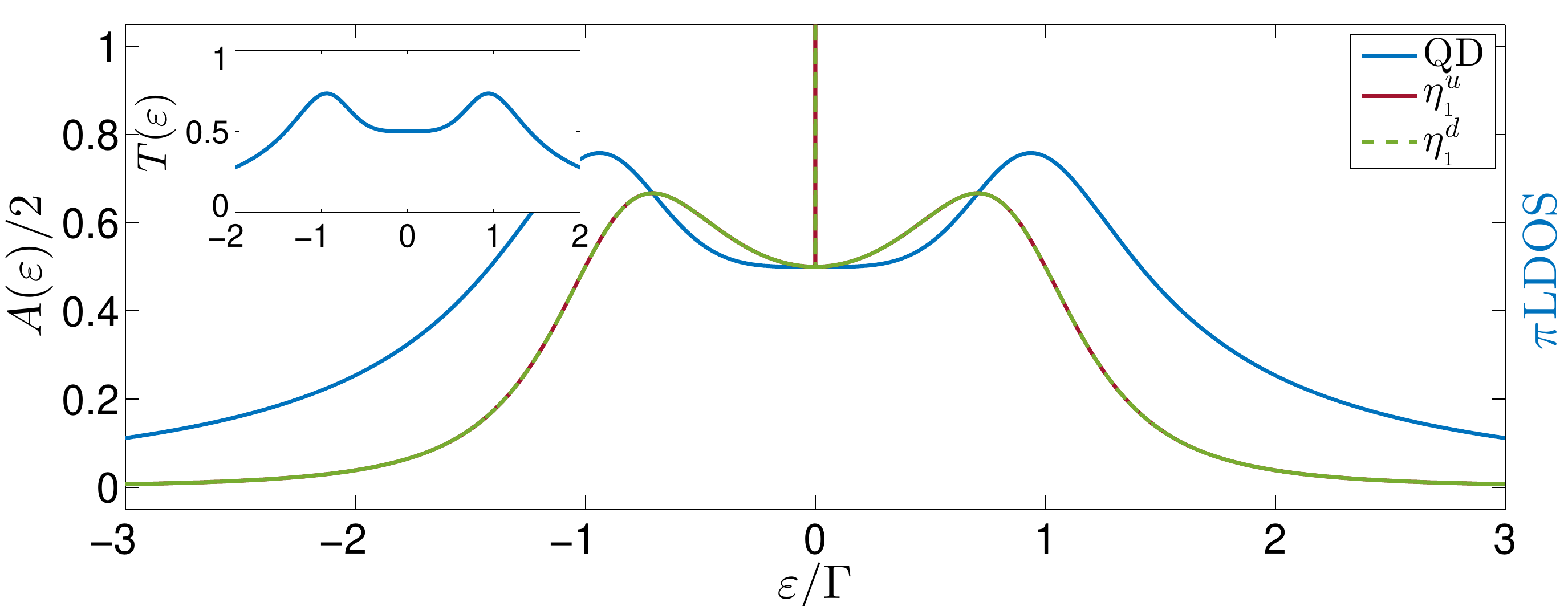}
\caption{Local density of states LDOS for QD (blue solid line) and spectral functions for MBSs $\eta_{1}^{u}$ (red solid line) and $\eta_{1}^{d}$ (green dashed line) as function of the energy. The first is proportional to $T(\varepsilon)$ as shown in the inset. Here $\varepsilon_{d}=\epsilon_{_M}^{u(d)}=0$.}
\label{fig.S1}
\end{figure}

\begin{figure}[!t]
\centering
\includegraphics[width=1.0\linewidth]{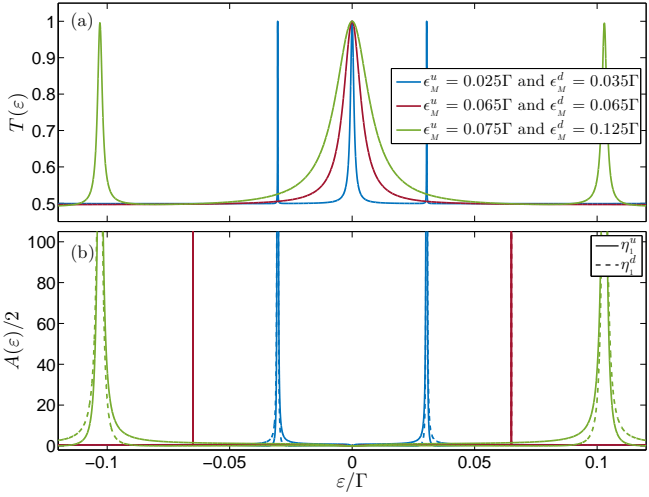}
\caption{Transmission $T(\varepsilon)$ through the QD (a) and spectral function $A(\varepsilon)$ for MBSs $\eta_{_M}^u$ and $\eta_{_M}^d$ (b) as function of energy. In both panels blue lines are for $\epsilon_{_M}^{d}=0.035\,\Gamma$ and $\epsilon_{_M}^{u}=0.025\,\Gamma$; red lines for $\epsilon_{_M}^{d(u)}=0.065\,\Gamma$; and green lines for $\epsilon_{_M}^{d}=0.125\,\Gamma$ and $\epsilon_{_M}^{u}=0.075\,\Gamma$. In panel (b) solid and dashed lines correspond to MBSs $\eta_{1}^{u}$ and $\eta_{1}^{d}$, respectively. Here $\varepsilon_{d}=0$. }
\label{fig.S2}
\end{figure}

First we consider the case with both TSCs wire lengths to be long enough to have vanishing coupling between $\eta_{_1}^{u(d)}$ and $\eta_{_2}^{u(d)}$, i. e. $\epsilon_{_M}^{u(d)} = 0$.  Figure\ \ref{fig.S1} shows LDOS for the QD, and spectral function for the MBSs $\eta_{1}^{d}$ and $\eta_{1}^{u}$. The LDOS (solid blue line) satisfies $T(\varepsilon)\propto\text{LDOS}(\varepsilon)$ according to Eqs.\ (\ref{tras}) and (\ref{LDOS}), and displays a half-maximum value at zero-energy, this being a MBS signature as it was proved by Liu \& and Baranger \cite{Liu2011}. The spectral function for MBSs coupled to the QD, $\eta_{_1}^{u}$ and $\eta_{_1}^{d}$ (red solid and green dashed lines, respectively), are strictly equivalent, both showing two symmetric wide resonances, placed at energy $\varepsilon=\pm\varepsilon_{2}$, due to the hybridization of the MBSs with the QD. Also, narrow peaks are observed pinned at zero-energy in the MBSs spectral function, as expected from vanishing inter MBSs coupling. The leakage of the latter states into QD is the responsibility of the Majorana behavior in transmission \cite{Vernek2014}, and in this case, as both TSCs have the same phase, they behave as an effective single TSC.

\begin{figure}[!t]
\centering
\includegraphics[width=1.0\linewidth]{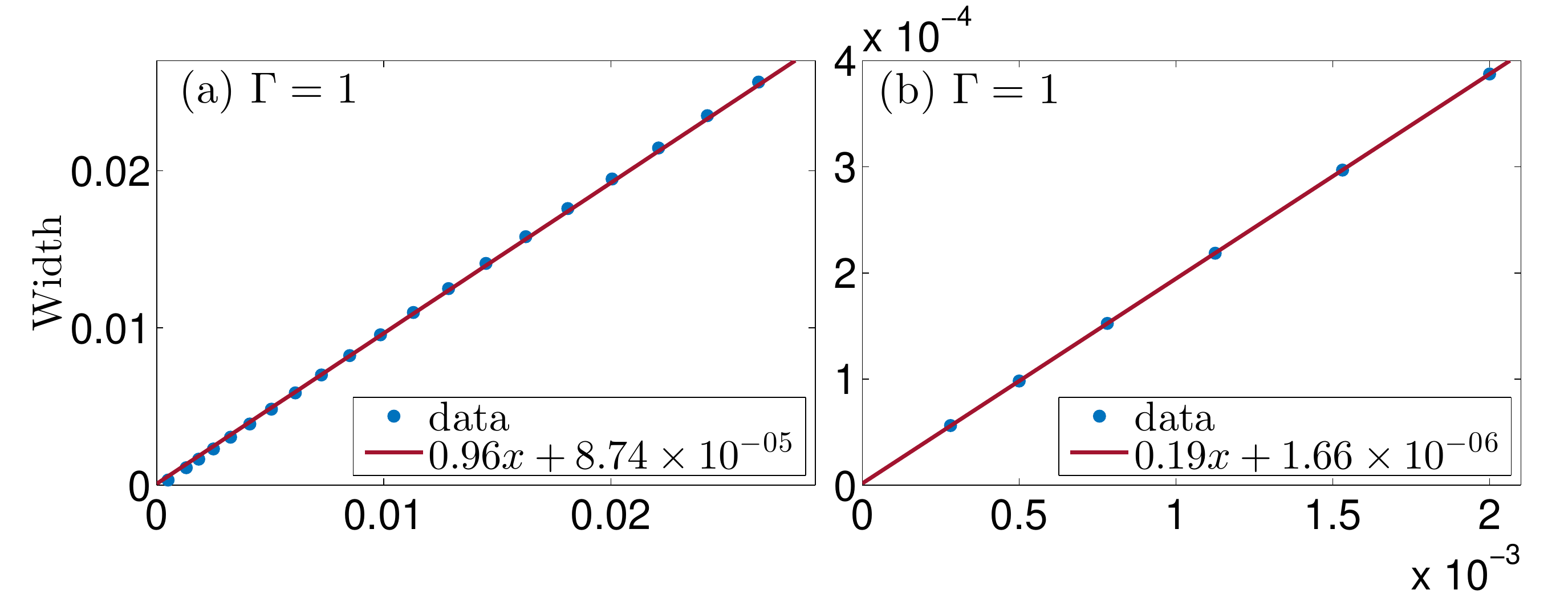}
\includegraphics[width=1.0\linewidth]{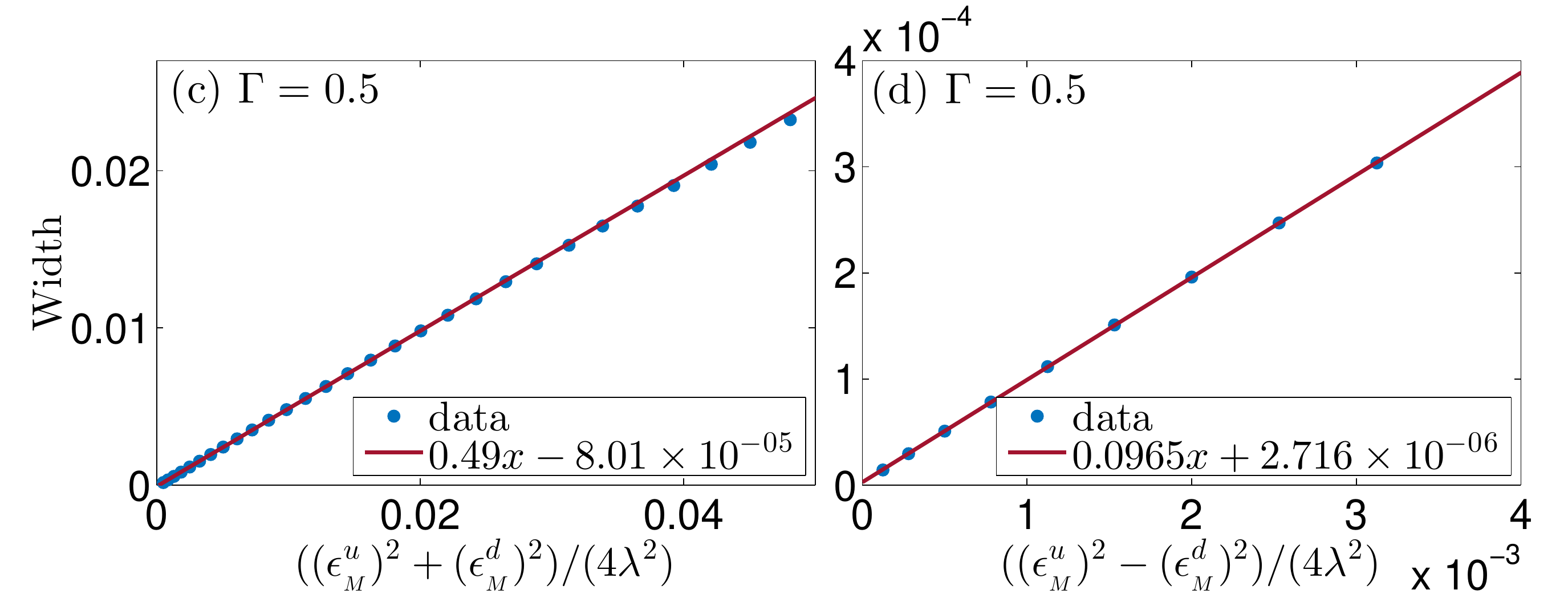}
\caption{[(a) and (c)] Half-height width of the central peak as function $\left(\epsilon_{_M}^{u}\right)^{2}+\left(\epsilon_{_M}^{d}\right)^{2}$ and [(b) and (d)] for lateral peaks of Fig.\ \ref{fig.S2}(a) as function $\left(\epsilon_{_M}^{u}\right)^{2}-\left(\epsilon_{_M}^{d}\right)^{2}$ for the same parameters of Fig.\ \ref{fig.S2}. As expected the slope in the four panels is linear with $\Gamma$.}
\label{fig.S3}
\end{figure}

Figure\ \ref{fig.S2} displays the transmission probability and the spectral function of the MBSs closely placed to the QD, where the coupling strength between $\eta_{_1}^{u(d)}$ and $\eta_{_2}^{u(d)}$ is included, by allowing $\epsilon_{_M}^{u(d)} \neq 0$. As we fixed small values of $\epsilon_{_M}^{u(d)}$ as compared to $\lambda_{u(d)}$; we focus on an energy region where this takes place. For the set of $\epsilon_{_M}^{u(d)}$ values used in Fig.\ \ref{fig.S2}, Eq.\ (\ref{E1}) takes the following values: $\pm0.03\,\Gamma$, $\pm0.065\,\Gamma$ and $\pm0.103\,\Gamma$, which are precisely the position of the peaks appearing in Fig.\ \ref{fig.S2}(b). It is worth to mention that these peaks have a projection on the transmission probability showed in Fig.\ \ref{fig.S2}(a) as sharp resonances, corresponding to quasi-BICs. They are located at the same energies mentioned above, except for the case with $\epsilon_{_M}^{d}=\epsilon_{_M}^{u}\neq 0$ (red lines), since these states become BICs. In this case, both peaks in the spectral function are strictly equivalent and since there is no phase difference, they interfere destructively and cancel each other. Then, both the transmission and the conductance do not have any signature of the BICs presence. This particular behavior allows us to assume that the width of the central peak (placed around $\varepsilon = 0$) is a function that depends on the sum of $\epsilon_{_M}^u$ and $\epsilon_{_M}^d$, while the width of the lateral peaks, given by Eq.\ (\ref{E1}), depends on the difference. It is supported by the fact that lateral peaks vanish when $\epsilon_{_M}^u = \epsilon_{_M}^d$, while the central peak remains. The symmetric side peaks observed in spectral function displayed on Fig.\ \ref{fig.S2}, have very subtle widths. Then, it follows that these could constitute bound states in the continuum. Following this, the transmission and spectral function in the TSC can be written approximately as
\begin{align}
&T(\varepsilon)\simeq\\
&\frac{1}{2}\left( 1+\frac{\gamma_{\text{c}}^{2}}{\varepsilon^{2}+\gamma_{\text{c}}^{2}}+\frac{\gamma_{\text{l},+}^{2}}{(\varepsilon-\varepsilon_{\text{l}}^{+})^2+\gamma_{\text{l},+}^2}+\frac{\gamma_{\text{l},-}^{2}}{(\varepsilon-\varepsilon_{\text{l}}^{-})^{2}+\gamma_{\text{l},-}^2} \right) \nonumber\,,
\end{align}
\begin{equation} A(\varepsilon)\simeq\frac{\gamma_{\text{l},+}}{(\varepsilon-\varepsilon_{\text{l}}^{+})^2+\gamma_{\text{l},+}^2}+\frac{\gamma_{\text{l},-}}{(\varepsilon-\varepsilon_{\text{l}}^{-})^{2}+\gamma_{\text{l},-}^2}\,,
\end{equation}
where $\gamma_{\text{c}}$ and $\gamma_{\text{l},+(-)}$ are the widths of the central and side peak located at the right(left), respectively, being
\begin{equation}\label{energyloc}
\varepsilon_{\text{l}}^{\pm}\simeq\sqrt{\frac{\left(\epsilon_{_M}^{u}\right)^{2}\pm\left(\epsilon_{_M}^{d}\right)^{2}}{2}}
\end{equation}
the peaks energy localization in the limit of weak inter MBSs coupling $\left(\epsilon_{_M}^{u(d)}\right)^{2}/\lambda^{2}\ll 1$. Central resonance satisfies $\gamma_{\text{c}}^{2}\propto\left(\epsilon_{_M}^{u}\right)^{2}+\left(\epsilon  _{_M}^{d}\right)^{2}$. On the other hand, setting $\epsilon_{_M}^{u(d)}=\epsilon_{_M}+(-)\Delta$ and considering $\Delta\ll\epsilon_{_M}$, we obtain $|\varepsilon_{\text{l}}|\sim\epsilon_{_M}(1-\Delta^{2}/\epsilon_{_M}^{2})$, then $\gamma_{\text{l}}\sim\Delta$. Therefore, in the case with $\Delta=0$, i. e. $\epsilon_{_M}^{u}=\epsilon_{_M}^{d}$, we have $\gamma_{\text{l}}=0$, thus the contribution from these states to the transmission vanishes, appearing as $\delta$-Dirac in the MBS spectral function. Accordingly, these states are essentially BICs. Besides, Fig.\ \ref{fig.S3} shows the peak width as a function of the parameter $\xi_{\pm}$, defined as $4\lambda^{2}\xi_{\pm}=(\epsilon_{_M}^u)^2 \pm ( \epsilon_{_M}^d)^2$, for central and lateral peaks in panels (a) and (b). We can observe that both peaks fulfill a linear dependence, the central one with $\xi_{_+}$ and the laterals with $\xi_{_-}$, verifying our expectation. For small values of $\xi_{_-}$  it can be approximated to $\Delta$.

\begin{figure}[!b]
\centering
\includegraphics[width=1.0\linewidth]{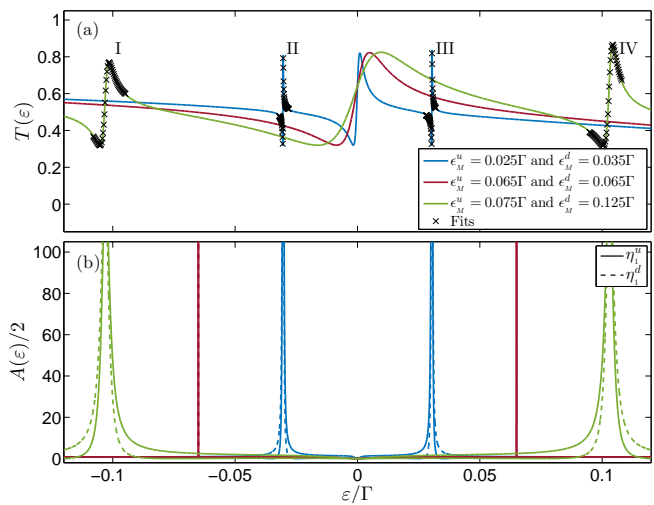}
\caption{Transmission $T$ (a) and MBSs spectral function (b) for fixed $\varepsilon_{d} = 0.75\,\Gamma$ as function of energy. In both TSCs, the MBSs have non-vanishing coupling, i.e. $\epsilon_{_M}^{u(d)}\neq 0$. Black crosses corresponds to the Fano-Majorana fitting given by Eq.\ (\ref{fano_fit}), which parameters are shown in Table\ \ref{tab.1}.}
\label{fig.S5}
\end{figure}

Furthermore, the signatures of the quasi-BICs in the transmission showed in Fig.\ \ref{fig.S2}, evolve to a different shape when the QD energy level is out of resonance ($\varepsilon_d \neq 0$). Using a fixed $\varepsilon_{d}=0.75\,\Gamma$, in Fig.\ \ref{fig.S5} we plot the transmission probability and spectral function for the MBSs $\eta_{1}^{u(d)}$ as function of energy. In Fig.\ \ref{fig.S5}(b) the BICs appear at energies described by Eq.\ (\ref{energyloc}). They still have a projection on the transmission as Fano-like shapes centered at the same energies, as we show in Fig.\ \ref{fig.S5}(a). But in contrast with Fig.\ \ref{fig.S2}, their amplitudes are modulated due to the occupancy in the QD, while the central peak also becomes a broad Fano line-shape. To characterize each Fano resonance in transmission due to the projection of BICs, we include its fitting with a modified general Fano line-shape expression, given by
\begin{equation}
  F(\varepsilon) = a\frac{| q \Gamma_{\text{eff}}/2 + \varepsilon - \varepsilon_{f}|^{2}}{\left(\Gamma_{\text{eff}}/2\right)^{2} + \left(\varepsilon - \varepsilon_{f}\right)^{2}}\text{.}
  \label{fano_fit}
\end{equation}
Here $q$ is the Fano factor ($q=q_{_r}+iq_{_i}$) and $a$ is the Fano-Majorana amplitude parameter. The values used for the parameters of these fits in Fig.\ \ref{fig.S5} are given in Table\ \ref{tab.1}. Note that $a$ has an approximated value of $1/2$ and that if $\epsilon_{_M}^{u} = \epsilon_{_M}^{d}$, again, we obtain BICs since the spectral function shows two Dirac-$\delta$ functions.
\begin{table}[!h]
    \begin{tabular}{ | c || c | c | c | c | c |}
    \hline
    $F(\varepsilon)$ & $q_{_r}$ & $q_{_i}$ & $\varepsilon_{f}$ & $\Gamma_{\text{eff}}$ & $a$      \\ \hline
    \hline
    I           & $0.4425$ & $0.990$  & $-0.1030$     & $0.003650$    & $0.5004$ \\ \hline
    II          & $0.4619$ & $1.016$  & $-0.0304$     & $0.000152$    & $0.5012$ \\ \hline
    III         & $0.4845$ & $1.041$  & $+0.0304$     & $0.000152$    & $0.4978$ \\ \hline
    IV          & $0.5174$ & $1.086$  & $+0.1031$     & $0.003581$    & $0.4849$ \\
    \hline
    \end{tabular}
    \caption{Parameters used by Eq.\ (\ref{fano_fit}) and shown as black crosses of Fig.\ \ref{fig.S5}. All energies are in units of $\Gamma$.}
    \label{tab.1}
\end{table}

\begin{figure}[!b]
\centering
\includegraphics[width=1.0\linewidth]{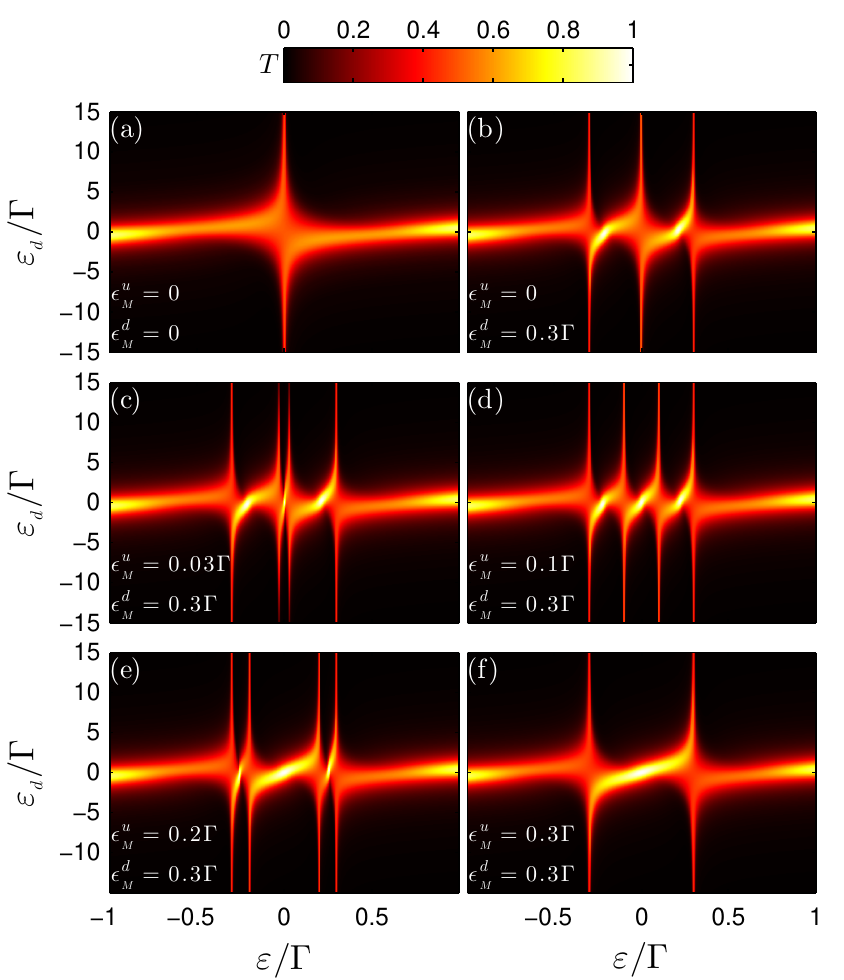}
\caption{Transmission $T(\varepsilon)$ contour plot of the single QD system as a function of the energy $\varepsilon$ and QD level $\varepsilon_{_0}$. Panel (a) $\epsilon_{_M}^u = \epsilon_{_M}^d = 0$. Panel (b) $\epsilon_{_M}^u = 0$ and $\epsilon_{_M}^d = 0.3\,\Gamma$. Panel (c) $\epsilon_{_M}^u = 0.03\,\Gamma$ and $\epsilon_{_M}^d = 0.3\,\Gamma$. Panel (d) $\epsilon_{_M}^u = 0.1\,\Gamma$ and $\epsilon_{_M}^d = 0.3\,\Gamma$. Panel (e) $\epsilon_{_M}^u = 0.2\,\Gamma$ and $\epsilon_{_M}^d = 0.3\,\Gamma$. Panel (f) $\epsilon_{_M}^u = 0.3\,\Gamma$ and $\epsilon_{_M}^d = 0.3\,\Gamma$.}
\label{fig.S6}
\end{figure}

When $\epsilon_{_M}^u = \epsilon_{_M}^d = 0$ and $\theta = 0$, Eqs.\ (\ref{E0}) and (\ref{E1}) take values $0$ (multiplicity 4) and Eq. (\ref{E2}) takes the values $\pm 2\lambda = \Gamma$ (multiplicity 2). In Fig.\ \ref{fig.S6}(a) we can see these eigenvalues represented as a $1/2$ transmission peak at zero-energy, regardless $\varepsilon_{_d}$. Besides, when $\varepsilon_{_d}=0$ and for energies around and greater than $\varepsilon = \Gamma$, the transmission probability takes its maximum value. This effect appears in all panels of Fig.\ \ref{fig.S6} and it is independent of the values of $\epsilon_{_M}^{u(d)}$. In Fig.\ \ref{fig.S6}(b) the Majorana coupling in each TSC takes the values $\epsilon_{_M}^u = 0$ and $\epsilon_{_M}^d = 0.3\,\Gamma$, when the QD is in resonance ($\varepsilon_{_d} = 0$) two sharp peaks appear at an approximated energy $\pm 0.21\,\Gamma$ given by Eqs.\ (\ref{E1}) and (\ref{E2}). Both peaks reach a maximum value of transmission near unity. As the QD is taken out of resonance ($\varepsilon_{_d} \neq 0$), these peaks evolve to a half-maximum value located at $\varepsilon=\pm\epsilon_{_M}^d = \pm 0.3\,\Gamma$. The zero-energy $1/2$ peak evolves in the same way as Fig.\ \ref{fig.S6}(a) due to that one of the TSC still have vanishing inter-MBSs coupling $\epsilon_{_M}^u = 0$. This constitutes a proof that splitted MBSs are also leaking into the QD. Now we allow $\epsilon_{_M}^u$ to take values different from zero but rather small as compare to others parameters. We can see in Fig.\ \ref{fig.S6}(c) how the central peak takes values near unity for $\varepsilon_{_d}=0$, and the off-resonance ($\varepsilon_{d}\neq0$) peaks split quickly into two, located at energies $\varepsilon=\pm\epsilon_{_M}^u = \pm 0.03\,\Gamma$. As the value of $\epsilon_{_M}^u$ increases [Fig.\ \ref{fig.S6}(d)], the central peaks at $\varepsilon = 0$ and $\varepsilon_{_d} = 0$ become wider and the two new off-resonance peaks shift to energies $\varepsilon=\pm\epsilon_{_M}^u = \pm 0.1\,\Gamma$. Figure\ \ref{fig.S6}(e) shows the same behavior as Fig.\ \ref{fig.S6}(d): note how the central peak spreads out while the lateral ones shrink. Finally, in Fig.\ \ref{fig.S6}(f) both TSCs have equal MBSs coupling $\epsilon_{_M}^u = \epsilon_{_M}^d$ and the lateral peaks fall to $1/2$ transmission.

\subsection{With phase difference, $\theta\neq 0$}

\begin{figure}[!b]
\centering
\includegraphics[width=1.0\linewidth]{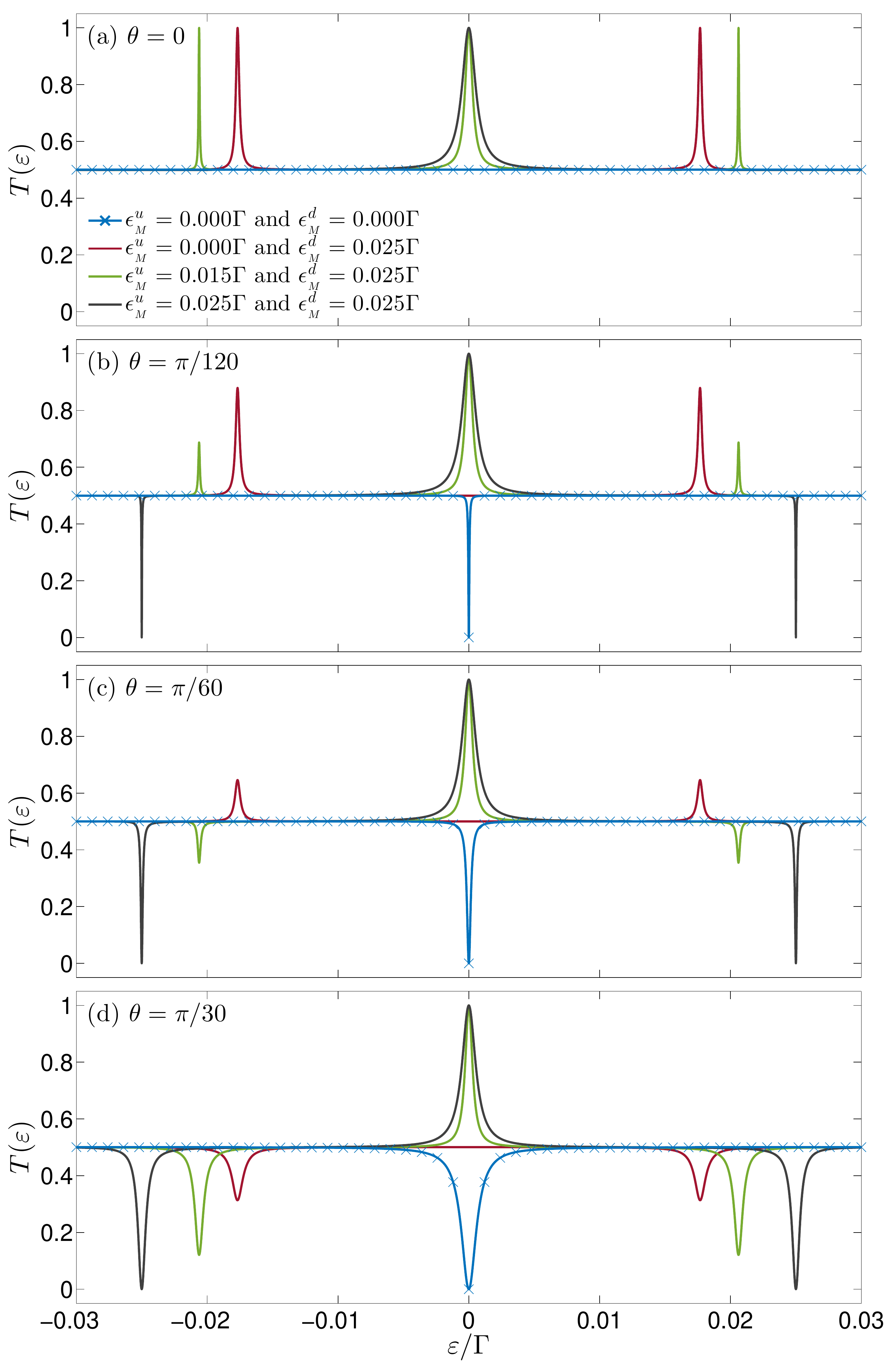}
\caption{Transmission $T$ as function of the energy for different combination of the inter-MBSs couplings $\epsilon_{_M}^{u}$ and $\epsilon_{_M}^{d}$. Different panels are displayed for small values of the phase angle: (a) $\theta=0$; (b) $\theta=\pi/120$, (c) $\theta=\pi/60$ and (d) $\theta=\pi/30$.}
\label{fig.S7}
\end{figure}

In this subsection, we consider the case with a general phase difference between both TSCs, to study the robustness of the above results. In Fig.\ \ref{fig.S7} we display the transmission through QD for different pairs of $\epsilon_{_M}^{d}$ and $\epsilon_{_M}^{u}$ for four fixed small ($\theta << 1$) phase difference.

\begin{figure}[!b]
\centering
\includegraphics[width=1.0\linewidth]{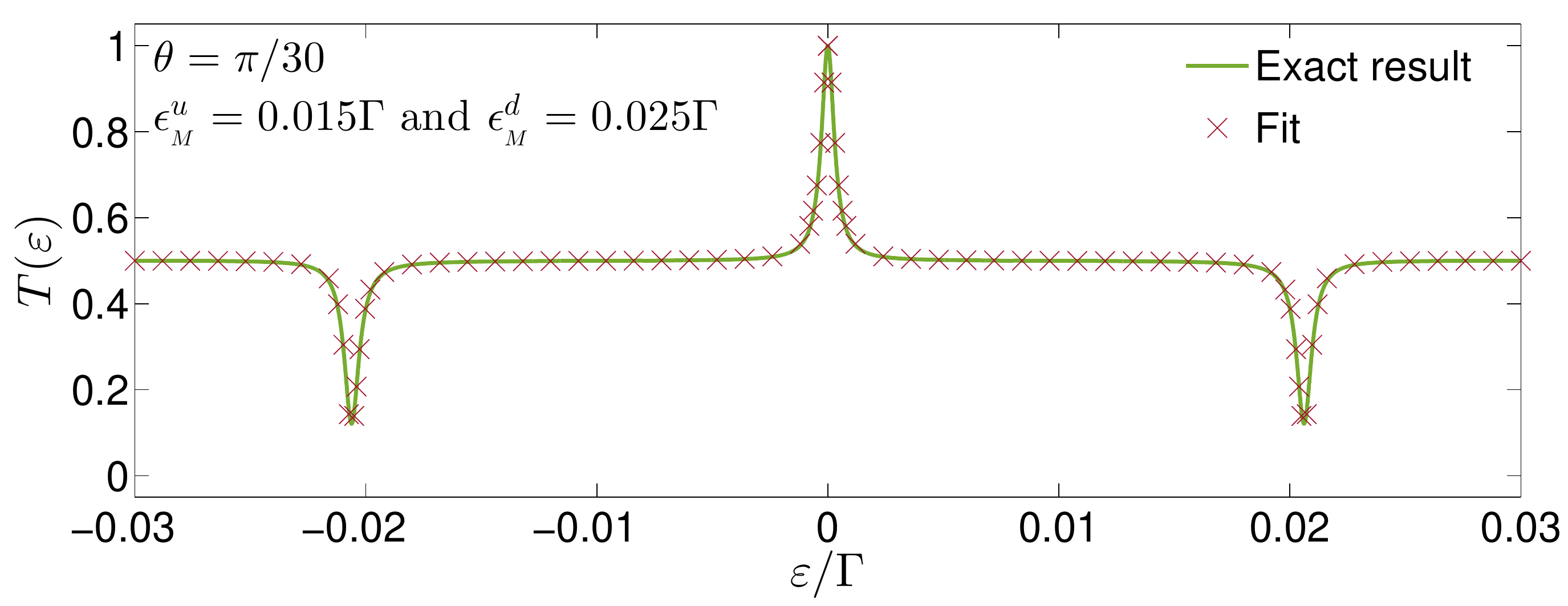}
\caption{Transmission $T$ as function of the energy. Here $\epsilon_{_M}^{u}=0.015\Gamma$ and $\epsilon_{_M}^{d}=0.025\Gamma$ and $\theta=\pi/30$. Solid lines corresponds to the exact result and crosses to the fit.}
\label{fig.S8}
\end{figure}

For the case $\epsilon_{_M}^{u(d)}=0$ (blue line and crosses) a destructive interference between both leaked MBSs into QD is observed whenever $\theta\neq 0$, obtaining a vanishing transmission at zero-energy. A similar effect occurs when both inter-Majorana couplings are equal but nonzero, $\epsilon_{_M}^{u}= \epsilon_{_M}^{d} = \epsilon_{_M} \neq 0$ (black lines), but in this case the vanishing transmissions are pinned at the values $\pm\epsilon_{_M}$. Furthermore, a maximum transmission is always obtained at $\varepsilon=\varepsilon_{d}=0$, independent of the phase difference. Then, the resonant state due to QD is entirely uncovered by MBSs phenomenology. It is worth to mention that the sharp anti-resonances obtained for both cases discussed above are symmetrical, and therefore each of them can be characterized through Fano line shape with imaginary $q$ values.

We consider inter-MBSs couplings such as $\epsilon_{_M}^{u}=0$ and $\epsilon_{_M}^{d}\neq 0$ (red lines) a half-maximum transmission is obtained at zero-energy regardless the phase difference, while sharp resonant states are pinned around energies $\varepsilon=\pm\epsilon_{_M}^{d}$. These states correspond to quasi-BICs projected into the transmission. For the case with $\epsilon_{_M}^{u} \neq \epsilon_{_M}^{d} \neq 0$ (green lines), a maximum transmission is obtained at zero-energy regardless the phase difference, same as black lines. However, each sharp lateral resonances shown for $\theta = 0$, evolves to an amplitude decreased resonance, and then to a non-vanishing anti-resonances as $\theta$ is moving away from 0. It is interesting to note that both resonances and/or anti-resonances are symmetrical regardless $\theta$, which are associated to predominant imaginary $q$ values ($q_{r}<<1$) for the Fano factor, as was mentioned above. In this scenario, we present a qualitative and analytic expression to characterize the behavior of the transmission, as a superposition of both Fano and Breit-Wigner line shapes,

\begin{align}\label{trasfano}
    &T(\varepsilon)\sim\\&\frac{1}{2}\left[\cos^{2}[\phi(\theta)]+\frac{1}{q_{i}^{2}+1}\frac{[\left(|\varepsilon| - \varepsilon_{f} \right)/\Gamma_f]^{2}+q_{i}^{2}}{[\left(|\varepsilon| - \varepsilon_{f} \right)/\Gamma_f]^{2}+1} + \frac{\Gamma_0^2}{\varepsilon^2+\Gamma_0^2}\right]\nonumber ,
\end{align}
where $\phi(\theta)$ is a function of the phase difference $\theta$, $\Gamma_f$ is the width of the lateral anti-resonances (resonances),$\Gamma_0$ is the width of the central resonance and the imaginary Fano factor is such as $q_{i}\propto\left( \left|\epsilon_{_M}^{d}-\epsilon_{_M}^{u}\right|/\Gamma\right)\cot{[\phi(\theta)]}$. For instance, for a specific case, Fig.\ \ref{fig.S8} displays a comparison between the exact result and the  fitting using Eq. (\ref{trasfano}). The equivalence is remarkable, and then it is clear that Eq. (\ref{trasfano}) describes the transmission behavior regarding the TSCs phase difference properly.

\section{Summary}\label{Summary}

We have studied the transmission across a QD coupled with two TSCs, embedded between two normal leads, used as a probe of MBSs interactions. For the case without phase difference ($\theta=0$), our results showed that the BICs projection in transmission arising from MBSs can be controlled by tuning the inter-MBSs coupling, as well as the modified Fano effect in the system, can be seen as a way to provide additional characterization. A possible application of the proposed setup can be achieved when the QD is far off resonance, in this case, the non-vanishing transmission is pinned around the inter-MBSs energies $\pm\epsilon_{_M}^{u(d)}$, as shown in Fig.\ \ref{fig.S6}. Then, the system can be considered to implement a calibrating device, allowing further characterization of TSCs hosting MBSs, used in the system proposed by the authors in \cite{Zambrano2018}. If the energies $\pm\epsilon_{_M}^{u(d)}$ are known, it is possible to determine the suitable manipulation of gate voltages, proposed in \cite{Alicea2011}, to readout the protected information stored in the BIC.

The authors acknowledge financial support from FONDECYT under Grant 1180914, CONICYT under Grant PAI-79140064 and UTFSM grant FI4060. J.P.R.-A is grateful for the funding of scholarship CONICYT-Chile No. 21141034 and PIIC-UTFSM grant.

\bibliography{biblio_majorana_20181130}

\begin{thebibliography}{60}%
\makeatletter
\providecommand \@ifxundefined [1]{%
 \@ifx{#1\undefined}
}%
\providecommand \@ifnum [1]{%
 \ifnum #1\expandafter \@firstoftwo
 \else \expandafter \@secondoftwo
 \fi
}%
\providecommand \@ifx [1]{%
 \ifx #1\expandafter \@firstoftwo
 \else \expandafter \@secondoftwo
 \fi
}%
\providecommand \natexlab [1]{#1}%
\providecommand \enquote  [1]{``#1''}%
\providecommand \bibnamefont  [1]{#1}%
\providecommand \bibfnamefont [1]{#1}%
\providecommand \citenamefont [1]{#1}%
\providecommand \href@noop [0]{\@secondoftwo}%
\providecommand \href [0]{\begingroup \@sanitize@url \@href}%
\providecommand \@href[1]{\@@startlink{#1}\@@href}%
\providecommand \@@href[1]{\endgroup#1\@@endlink}%
\providecommand \@sanitize@url [0]{\catcode `\\12\catcode `\$12\catcode
  `\&12\catcode `\#12\catcode `\^12\catcode `\_12\catcode `\%12\relax}%
\providecommand \@@startlink[1]{}%
\providecommand \@@endlink[0]{}%
\providecommand \url  [0]{\begingroup\@sanitize@url \@url }%
\providecommand \@url [1]{\endgroup\@href {#1}{\urlprefix }}%
\providecommand \urlprefix  [0]{URL }%
\providecommand \Eprint [0]{\href }%
\providecommand \doibase [0]{http://dx.doi.org/}%
\providecommand \selectlanguage [0]{\@gobble}%
\providecommand \bibinfo  [0]{\@secondoftwo}%
\providecommand \bibfield  [0]{\@secondoftwo}%
\providecommand \translation [1]{[#1]}%
\providecommand \BibitemOpen [0]{}%
\providecommand \bibitemStop [0]{}%
\providecommand \bibitemNoStop [0]{.\EOS\space}%
\providecommand \EOS [0]{\spacefactor3000\relax}%
\providecommand \BibitemShut  [1]{\csname bibitem#1\endcsname}%
\let\auto@bib@innerbib\@empty
\bibitem [{\citenamefont {Majorana}(1937)}]{Majorana1937}%
  \BibitemOpen
  \bibfield  {author} {\bibinfo {author} {\bibfnamefont {E.}~\bibnamefont
  {Majorana}},\ }\href {\doibase 10.1007/BF02961314} {\bibfield  {journal}
  {\bibinfo  {journal} {Nuovo Cimento}\ }\textbf {\bibinfo {volume} {14}},\
  \bibinfo {pages} {171} (\bibinfo {year} {1937})}\BibitemShut {NoStop}%
\bibitem [{\citenamefont {Wilczek}(2009)}]{Wilczek2009}%
  \BibitemOpen
  \bibfield  {author} {\bibinfo {author} {\bibfnamefont {F.}~\bibnamefont
  {Wilczek}},\ }\href {\doibase 10.1038/nphys1380} {\bibfield  {journal}
  {\bibinfo  {journal} {Nat. Phys.}\ }\textbf {\bibinfo {volume} {5}},\
  \bibinfo {pages} {614} (\bibinfo {year} {2009})}\BibitemShut {NoStop}%
\bibitem [{\citenamefont {Franz}(2010)}]{Franz2010}%
  \BibitemOpen
  \bibfield  {author} {\bibinfo {author} {\bibfnamefont {M.}~\bibnamefont
  {Franz}},\ }\href {https://physics.aps.org/articles/v3/24} {\bibfield
  {journal} {\bibinfo  {journal} {Physics}\ }\textbf {\bibinfo {volume} {3}},\
  \bibinfo {pages} {24} (\bibinfo {year} {2010})}\BibitemShut {NoStop}%
\bibitem [{\citenamefont {Wu}\ and\ \citenamefont {Cao}(2012)}]{Wu2012}%
  \BibitemOpen
  \bibfield  {author} {\bibinfo {author} {\bibfnamefont {B.~H.}\ \bibnamefont
  {Wu}}\ and\ \bibinfo {author} {\bibfnamefont {J.~C.}\ \bibnamefont {Cao}},\
  }\href {\doibase 10.1103/PhysRevB.85.085415} {\bibfield  {journal} {\bibinfo
  {journal} {Phys. Rev. B}\ }\textbf {\bibinfo {volume} {85}},\ \bibinfo
  {pages} {085415} (\bibinfo {year} {2012})}\BibitemShut {NoStop}%
\bibitem [{\citenamefont {Nayak}\ \emph {et~al.}(2008)\citenamefont {Nayak},
  \citenamefont {Simon}, \citenamefont {Stern}, \citenamefont {Freedman},\ and\
  \citenamefont {Das~Sarma}}]{Nayak2008}%
  \BibitemOpen
  \bibfield  {author} {\bibinfo {author} {\bibfnamefont {C.}~\bibnamefont
  {Nayak}}, \bibinfo {author} {\bibfnamefont {S.~H.}\ \bibnamefont {Simon}},
  \bibinfo {author} {\bibfnamefont {A.}~\bibnamefont {Stern}}, \bibinfo
  {author} {\bibfnamefont {M.}~\bibnamefont {Freedman}}, \ and\ \bibinfo
  {author} {\bibfnamefont {S.}~\bibnamefont {Das~Sarma}},\ }\href {\doibase
  10.1103/RevModPhys.80.1083} {\bibfield  {journal} {\bibinfo  {journal} {Rev.
  Mod. Phys.}\ }\textbf {\bibinfo {volume} {80}},\ \bibinfo {pages} {1083}
  (\bibinfo {year} {2008})}\BibitemShut {NoStop}%
\bibitem [{\citenamefont {Leijnse}\ and\ \citenamefont
  {Flensberg}(2011)}]{Leijnse2011}%
  \BibitemOpen
  \bibfield  {author} {\bibinfo {author} {\bibfnamefont {M.}~\bibnamefont
  {Leijnse}}\ and\ \bibinfo {author} {\bibfnamefont {K.}~\bibnamefont
  {Flensberg}},\ }\href {\doibase 10.1103/PhysRevLett.107.210502} {\bibfield
  {journal} {\bibinfo  {journal} {Phys. Rev. Lett.}\ }\textbf {\bibinfo
  {volume} {107}},\ \bibinfo {pages} {210502} (\bibinfo {year}
  {2011})}\BibitemShut {NoStop}%
\bibitem [{\citenamefont {Bravyi}\ and\ \citenamefont
  {Kitaev}(2002)}]{Bravyi2000}%
  \BibitemOpen
  \bibfield  {author} {\bibinfo {author} {\bibfnamefont {S.~B.}\ \bibnamefont
  {Bravyi}}\ and\ \bibinfo {author} {\bibfnamefont {A.~Y.}\ \bibnamefont
  {Kitaev}},\ }\href
  {http://www.sciencedirect.com/science/article/pii/S0003491602962548}
  {\bibfield  {journal} {\bibinfo  {journal} {Ann. Phys.}\ }\textbf {\bibinfo
  {volume} {298}},\ \bibinfo {pages} {210 } (\bibinfo {year}
  {2002})}\BibitemShut {NoStop}%
\bibitem [{\citenamefont {Kitaev}(2001)}]{Kitaev2001}%
  \BibitemOpen
  \bibfield  {author} {\bibinfo {author} {\bibfnamefont {A.~Y.}\ \bibnamefont
  {Kitaev}},\ }\href {http://stacks.iop.org/1063-7869/44/i=10S/a=S29}
  {\bibfield  {journal} {\bibinfo  {journal} {Phys. Usp.}\ }\textbf {\bibinfo
  {volume} {44}},\ \bibinfo {pages} {131} (\bibinfo {year} {2001})}\BibitemShut
  {NoStop}%
\bibitem [{\citenamefont {Kitaev}(2003)}]{Kitaev2003}%
  \BibitemOpen
  \bibfield  {author} {\bibinfo {author} {\bibfnamefont {A.}~\bibnamefont
  {Kitaev}},\ }\href
  {http://www.sciencedirect.com/science/article/pii/S0003491602000180}
  {\bibfield  {journal} {\bibinfo  {journal} {Ann. Phys.}\ }\textbf {\bibinfo
  {volume} {303}},\ \bibinfo {pages} {2 } (\bibinfo {year} {2003})}\BibitemShut
  {NoStop}%
\bibitem [{\citenamefont {Pachos}(2012)}]{Pachos2012}%
  \BibitemOpen
  \bibfield  {author} {\bibinfo {author} {\bibfnamefont {J.~K.}\ \bibnamefont
  {Pachos}},\ }\href
  {http://www.cambridge.org/cl/academic/subjects/physics/quantum-physics-quantum-information-and-quantum-computation/introduction-topological-quantum-computation?format=HB&isbn=9781107005044#kTKO0fLo9lhv1Li6.97}
  {\emph {\bibinfo {title} {Introduction to Topological Quantum
  Computation}}},\ \bibinfo {edition} {1st}\ ed.\ (\bibinfo  {publisher}
  {Cambridge University Press},\ \bibinfo {address} {New York, NY, USA},\
  \bibinfo {year} {2012})\BibitemShut {NoStop}%
\bibitem [{\citenamefont {Kraus}\ \emph {et~al.}(2013)\citenamefont {Kraus},
  \citenamefont {Dalmonte}, \citenamefont {Baranov}, \citenamefont
  {L\"auchli},\ and\ \citenamefont {Zoller}}]{Kraus2013}%
  \BibitemOpen
  \bibfield  {author} {\bibinfo {author} {\bibfnamefont {C.~V.}\ \bibnamefont
  {Kraus}}, \bibinfo {author} {\bibfnamefont {M.}~\bibnamefont {Dalmonte}},
  \bibinfo {author} {\bibfnamefont {M.~A.}\ \bibnamefont {Baranov}}, \bibinfo
  {author} {\bibfnamefont {A.~M.}\ \bibnamefont {L\"auchli}}, \ and\ \bibinfo
  {author} {\bibfnamefont {P.}~\bibnamefont {Zoller}},\ }\href {\doibase
  10.1103/PhysRevLett.111.173004} {\bibfield  {journal} {\bibinfo  {journal}
  {Phys. Rev. Lett.}\ }\textbf {\bibinfo {volume} {111}},\ \bibinfo {pages}
  {173004} (\bibinfo {year} {2013})}\BibitemShut {NoStop}%
\bibitem [{\citenamefont {Albrecht}\ \emph {et~al.}(2016)\citenamefont
  {Albrecht}, \citenamefont {Higginbotham}, \citenamefont {Madsen},
  \citenamefont {Kuemmeth}, \citenamefont {Jespersen}, \citenamefont
  {Nyg{\aa}rd}, \citenamefont {Krogstrup},\ and\ \citenamefont
  {Marcus}}]{Albrecht2016}%
  \BibitemOpen
  \bibfield  {author} {\bibinfo {author} {\bibfnamefont {S.}~\bibnamefont
  {Albrecht}}, \bibinfo {author} {\bibfnamefont {A.}~\bibnamefont
  {Higginbotham}}, \bibinfo {author} {\bibfnamefont {M.}~\bibnamefont
  {Madsen}}, \bibinfo {author} {\bibfnamefont {F.}~\bibnamefont {Kuemmeth}},
  \bibinfo {author} {\bibfnamefont {T.}~\bibnamefont {Jespersen}}, \bibinfo
  {author} {\bibfnamefont {J.}~\bibnamefont {Nyg{\aa}rd}}, \bibinfo {author}
  {\bibfnamefont {P.}~\bibnamefont {Krogstrup}}, \ and\ \bibinfo {author}
  {\bibfnamefont {C.}~\bibnamefont {Marcus}},\ }\href
  {http://www.nature.com/nature/journal/v531/n7593/full/nature17162.html?WT.feed_name=subjects_electronics-photonics-and-device-physics}
  {\bibfield  {journal} {\bibinfo  {journal} {Nature}\ }\textbf {\bibinfo
  {volume} {531}},\ \bibinfo {pages} {206} (\bibinfo {year}
  {2016})}\BibitemShut {NoStop}%
\bibitem [{\citenamefont {Beenakker}(2013)}]{Beenakker2013}%
  \BibitemOpen
  \bibfield  {author} {\bibinfo {author} {\bibfnamefont {C.}~\bibnamefont
  {Beenakker}},\ }\href {\doibase 10.1146/annurev-conmatphys-030212-184337}
  {\bibfield  {journal} {\bibinfo  {journal} {Annu. Rev. Condens. Matter
  Phys.}\ }\textbf {\bibinfo {volume} {4}},\ \bibinfo {pages} {113} (\bibinfo
  {year} {2013})}\BibitemShut {NoStop}%
\bibitem [{\citenamefont {Laflamme}\ \emph {et~al.}(2014)\citenamefont
  {Laflamme}, \citenamefont {Baranov}, \citenamefont {Zoller},\ and\
  \citenamefont {Kraus}}]{Laflamme2014}%
  \BibitemOpen
  \bibfield  {author} {\bibinfo {author} {\bibfnamefont {C.}~\bibnamefont
  {Laflamme}}, \bibinfo {author} {\bibfnamefont {M.~A.}\ \bibnamefont
  {Baranov}}, \bibinfo {author} {\bibfnamefont {P.}~\bibnamefont {Zoller}}, \
  and\ \bibinfo {author} {\bibfnamefont {C.~V.}\ \bibnamefont {Kraus}},\ }\href
  {\doibase 10.1103/PhysRevA.89.022319} {\bibfield  {journal} {\bibinfo
  {journal} {Phys. Rev. A}\ }\textbf {\bibinfo {volume} {89}},\ \bibinfo
  {pages} {022319} (\bibinfo {year} {2014})}\BibitemShut {NoStop}%
\bibitem [{\citenamefont {Moore}(2009)}]{Moore2009}%
  \BibitemOpen
  \bibfield  {author} {\bibinfo {author} {\bibfnamefont {J.}~\bibnamefont
  {Moore}},\ }\href
  {https://www.nature.com/nphys/journal/v5/n6/full/nphys1294.html} {\bibfield
  {journal} {\bibinfo  {journal} {Nat. Phys.}\ }\textbf {\bibinfo {volume}
  {5}},\ \bibinfo {pages} {378} (\bibinfo {year} {2009})}\BibitemShut {NoStop}%
\bibitem [{\citenamefont {Semenoff}\ and\ \citenamefont
  {Sodano}(2006)}]{Semenoff2006}%
  \BibitemOpen
  \bibfield  {author} {\bibinfo {author} {\bibfnamefont {G.~W.}\ \bibnamefont
  {Semenoff}}\ and\ \bibinfo {author} {\bibfnamefont {P.}~\bibnamefont
  {Sodano}},\ }\href {https://arxiv.org/abs/cond-mat/0601261} {\bibfield
  {journal} {\bibinfo  {journal} {arXiv preprint cond-mat/0601261}\ } (\bibinfo
  {year} {2006})}\BibitemShut {NoStop}%
\bibitem [{\citenamefont {Tewari}\ \emph {et~al.}(2008)\citenamefont {Tewari},
  \citenamefont {Zhang}, \citenamefont {Das~Sarma}, \citenamefont {Nayak},\
  and\ \citenamefont {Lee}}]{Tewari2008}%
  \BibitemOpen
  \bibfield  {author} {\bibinfo {author} {\bibfnamefont {S.}~\bibnamefont
  {Tewari}}, \bibinfo {author} {\bibfnamefont {C.}~\bibnamefont {Zhang}},
  \bibinfo {author} {\bibfnamefont {S.}~\bibnamefont {Das~Sarma}}, \bibinfo
  {author} {\bibfnamefont {C.}~\bibnamefont {Nayak}}, \ and\ \bibinfo {author}
  {\bibfnamefont {D.-H.}\ \bibnamefont {Lee}},\ }\href {\doibase
  10.1103/PhysRevLett.100.027001} {\bibfield  {journal} {\bibinfo  {journal}
  {Phys. Rev. Lett.}\ }\textbf {\bibinfo {volume} {100}},\ \bibinfo {pages}
  {027001} (\bibinfo {year} {2008})}\BibitemShut {NoStop}%
\bibitem [{\citenamefont {Law}\ \emph {et~al.}(2009)\citenamefont {Law},
  \citenamefont {Lee},\ and\ \citenamefont {Ng}}]{Law2009}%
  \BibitemOpen
  \bibfield  {author} {\bibinfo {author} {\bibfnamefont {K.~T.}\ \bibnamefont
  {Law}}, \bibinfo {author} {\bibfnamefont {P.~A.}\ \bibnamefont {Lee}}, \ and\
  \bibinfo {author} {\bibfnamefont {T.~K.}\ \bibnamefont {Ng}},\ }\href
  {\doibase 10.1103/PhysRevLett.103.237001} {\bibfield  {journal} {\bibinfo
  {journal} {Phys. Rev. Lett.}\ }\textbf {\bibinfo {volume} {103}},\ \bibinfo
  {pages} {237001} (\bibinfo {year} {2009})}\BibitemShut {NoStop}%
\bibitem [{\citenamefont {Pikulin}\ \emph {et~al.}(2012)\citenamefont
  {Pikulin}, \citenamefont {Dahlhaus}, \citenamefont {Wimmer}, \citenamefont
  {Schomerus},\ and\ \citenamefont {Beenakker}}]{Pikulin2012}%
  \BibitemOpen
  \bibfield  {author} {\bibinfo {author} {\bibfnamefont {D.~I.}\ \bibnamefont
  {Pikulin}}, \bibinfo {author} {\bibfnamefont {J.~P.}\ \bibnamefont
  {Dahlhaus}}, \bibinfo {author} {\bibfnamefont {M.}~\bibnamefont {Wimmer}},
  \bibinfo {author} {\bibfnamefont {H.}~\bibnamefont {Schomerus}}, \ and\
  \bibinfo {author} {\bibfnamefont {C.~W.~J.}\ \bibnamefont {Beenakker}},\
  }\href {http://stacks.iop.org/1367-2630/14/i=12/a=125011} {\bibfield
  {journal} {\bibinfo  {journal} {New J. Phys.}\ }\textbf {\bibinfo {volume}
  {14}},\ \bibinfo {pages} {125011} (\bibinfo {year} {2012})}\BibitemShut
  {NoStop}%
\bibitem [{\citenamefont {Franz}(2013)}]{Franz2013}%
  \BibitemOpen
  \bibfield  {author} {\bibinfo {author} {\bibfnamefont {M.}~\bibnamefont
  {Franz}},\ }\href
  {http://www.nature.com/nnano/journal/v8/n3/full/nnano.2013.33.html?foxtrotcallback=true}
  {\bibfield  {journal} {\bibinfo  {journal} {Nat. nanotech.}\ }\textbf
  {\bibinfo {volume} {8}},\ \bibinfo {pages} {149} (\bibinfo {year}
  {2013})}\BibitemShut {NoStop}%
\bibitem [{\citenamefont {Prada}\ \emph {et~al.}(2012)\citenamefont {Prada},
  \citenamefont {San-Jose},\ and\ \citenamefont {Aguado}}]{Prada2012}%
  \BibitemOpen
  \bibfield  {author} {\bibinfo {author} {\bibfnamefont {E.}~\bibnamefont
  {Prada}}, \bibinfo {author} {\bibfnamefont {P.}~\bibnamefont {San-Jose}}, \
  and\ \bibinfo {author} {\bibfnamefont {R.}~\bibnamefont {Aguado}},\ }\href
  {\doibase 10.1103/PhysRevB.86.180503} {\bibfield  {journal} {\bibinfo
  {journal} {Phys. Rev. B}\ }\textbf {\bibinfo {volume} {86}},\ \bibinfo
  {pages} {180503(R)} (\bibinfo {year} {2012})}\BibitemShut {NoStop}%
\bibitem [{\citenamefont {Rainis}\ \emph {et~al.}(2013)\citenamefont {Rainis},
  \citenamefont {Trifunovic}, \citenamefont {Klinovaja},\ and\ \citenamefont
  {Loss}}]{Rainis2013}%
  \BibitemOpen
  \bibfield  {author} {\bibinfo {author} {\bibfnamefont {D.}~\bibnamefont
  {Rainis}}, \bibinfo {author} {\bibfnamefont {L.}~\bibnamefont {Trifunovic}},
  \bibinfo {author} {\bibfnamefont {J.}~\bibnamefont {Klinovaja}}, \ and\
  \bibinfo {author} {\bibfnamefont {D.}~\bibnamefont {Loss}},\ }\href {\doibase
  10.1103/PhysRevB.87.024515} {\bibfield  {journal} {\bibinfo  {journal} {Phys.
  Rev. B}\ }\textbf {\bibinfo {volume} {87}},\ \bibinfo {pages} {024515}
  (\bibinfo {year} {2013})}\BibitemShut {NoStop}%
\bibitem [{\citenamefont {Cook}\ \emph {et~al.}(2012)\citenamefont {Cook},
  \citenamefont {Vazifeh},\ and\ \citenamefont {Franz}}]{Cook2012}%
  \BibitemOpen
  \bibfield  {author} {\bibinfo {author} {\bibfnamefont {A.~M.}\ \bibnamefont
  {Cook}}, \bibinfo {author} {\bibfnamefont {M.~M.}\ \bibnamefont {Vazifeh}}, \
  and\ \bibinfo {author} {\bibfnamefont {M.}~\bibnamefont {Franz}},\ }\href
  {\doibase 10.1103/PhysRevB.86.155431} {\bibfield  {journal} {\bibinfo
  {journal} {Phys. Rev. B}\ }\textbf {\bibinfo {volume} {86}},\ \bibinfo
  {pages} {155431} (\bibinfo {year} {2012})}\BibitemShut {NoStop}%
\bibitem [{\citenamefont {Liu}\ and\ \citenamefont {Lobos}(2013)}]{Liu2013}%
  \BibitemOpen
  \bibfield  {author} {\bibinfo {author} {\bibfnamefont {X.-J.}\ \bibnamefont
  {Liu}}\ and\ \bibinfo {author} {\bibfnamefont {A.~M.}\ \bibnamefont
  {Lobos}},\ }\href {\doibase 10.1103/PhysRevB.87.060504} {\bibfield  {journal}
  {\bibinfo  {journal} {Phys. Rev. B}\ }\textbf {\bibinfo {volume} {87}},\
  \bibinfo {pages} {060504(R)} (\bibinfo {year} {2013})}\BibitemShut {NoStop}%
\bibitem [{\citenamefont {Stanescu}\ \emph {et~al.}(2011)\citenamefont
  {Stanescu}, \citenamefont {Lutchyn},\ and\ \citenamefont
  {Das~Sarma}}]{Stanescu2011}%
  \BibitemOpen
  \bibfield  {author} {\bibinfo {author} {\bibfnamefont {T.~D.}\ \bibnamefont
  {Stanescu}}, \bibinfo {author} {\bibfnamefont {R.~M.}\ \bibnamefont
  {Lutchyn}}, \ and\ \bibinfo {author} {\bibfnamefont {S.}~\bibnamefont
  {Das~Sarma}},\ }\href {\doibase 10.1103/PhysRevB.84.144522} {\bibfield
  {journal} {\bibinfo  {journal} {Phys. Rev. B}\ }\textbf {\bibinfo {volume}
  {84}},\ \bibinfo {pages} {144522} (\bibinfo {year} {2011})}\BibitemShut
  {NoStop}%
\bibitem [{\citenamefont {Lee}\ \emph {et~al.}(2014)\citenamefont {Lee},
  \citenamefont {Jiang}, \citenamefont {Houzet}, \citenamefont {Aguado},
  \citenamefont {Lieber},\ and\ \citenamefont {De~Franceschi}}]{Lee2014}%
  \BibitemOpen
  \bibfield  {author} {\bibinfo {author} {\bibfnamefont {E.~J.}\ \bibnamefont
  {Lee}}, \bibinfo {author} {\bibfnamefont {X.}~\bibnamefont {Jiang}}, \bibinfo
  {author} {\bibfnamefont {M.}~\bibnamefont {Houzet}}, \bibinfo {author}
  {\bibfnamefont {R.}~\bibnamefont {Aguado}}, \bibinfo {author} {\bibfnamefont
  {C.~M.}\ \bibnamefont {Lieber}}, \ and\ \bibinfo {author} {\bibfnamefont
  {S.}~\bibnamefont {De~Franceschi}},\ }\href
  {http://www.nature.com/nnano/journal/v9/n1/full/nnano.2013.267.html}
  {\bibfield  {journal} {\bibinfo  {journal} {Nat. nanotech.}\ }\textbf
  {\bibinfo {volume} {9}},\ \bibinfo {pages} {79} (\bibinfo {year}
  {2014})}\BibitemShut {NoStop}%
\bibitem [{\citenamefont {Flensberg}(2010)}]{Flensberg2010}%
  \BibitemOpen
  \bibfield  {author} {\bibinfo {author} {\bibfnamefont {K.}~\bibnamefont
  {Flensberg}},\ }\href {\doibase 10.1103/PhysRevB.82.180516} {\bibfield
  {journal} {\bibinfo  {journal} {Phys. Rev. B}\ }\textbf {\bibinfo {volume}
  {82}},\ \bibinfo {pages} {180516(R)} (\bibinfo {year} {2010})}\BibitemShut
  {NoStop}%
\bibitem [{\citenamefont {Wimmer}\ \emph {et~al.}(2011)\citenamefont {Wimmer},
  \citenamefont {Akhmerov}, \citenamefont {Dahlhaus},\ and\ \citenamefont
  {Beenakker}}]{Wimmer2011}%
  \BibitemOpen
  \bibfield  {author} {\bibinfo {author} {\bibfnamefont {M.}~\bibnamefont
  {Wimmer}}, \bibinfo {author} {\bibfnamefont {A.~R.}\ \bibnamefont
  {Akhmerov}}, \bibinfo {author} {\bibfnamefont {J.~P.}\ \bibnamefont
  {Dahlhaus}}, \ and\ \bibinfo {author} {\bibfnamefont {C.~W.~J.}\ \bibnamefont
  {Beenakker}},\ }\href {http://stacks.iop.org/1367-2630/13/i=5/a=053016}
  {\bibfield  {journal} {\bibinfo  {journal} {New J. Phys.}\ }\textbf {\bibinfo
  {volume} {13}},\ \bibinfo {pages} {053016} (\bibinfo {year}
  {2011})}\BibitemShut {NoStop}%
\bibitem [{\citenamefont {Nilsson}\ \emph {et~al.}(2008)\citenamefont
  {Nilsson}, \citenamefont {Akhmerov},\ and\ \citenamefont
  {Beenakker}}]{Nilsson2008}%
  \BibitemOpen
  \bibfield  {author} {\bibinfo {author} {\bibfnamefont {J.}~\bibnamefont
  {Nilsson}}, \bibinfo {author} {\bibfnamefont {A.~R.}\ \bibnamefont
  {Akhmerov}}, \ and\ \bibinfo {author} {\bibfnamefont {C.~W.~J.}\ \bibnamefont
  {Beenakker}},\ }\href {\doibase 10.1103/PhysRevLett.101.120403} {\bibfield
  {journal} {\bibinfo  {journal} {Phys. Rev. Lett.}\ }\textbf {\bibinfo
  {volume} {101}},\ \bibinfo {pages} {120403} (\bibinfo {year}
  {2008})}\BibitemShut {NoStop}%
\bibitem [{\citenamefont {Bolech}\ and\ \citenamefont
  {Demler}(2007)}]{Bolech2007}%
  \BibitemOpen
  \bibfield  {author} {\bibinfo {author} {\bibfnamefont {C.~J.}\ \bibnamefont
  {Bolech}}\ and\ \bibinfo {author} {\bibfnamefont {E.}~\bibnamefont
  {Demler}},\ }\href {\doibase 10.1103/PhysRevLett.98.237002} {\bibfield
  {journal} {\bibinfo  {journal} {Phys. Rev. Lett.}\ }\textbf {\bibinfo
  {volume} {98}},\ \bibinfo {pages} {237002} (\bibinfo {year}
  {2007})}\BibitemShut {NoStop}%
\bibitem [{\citenamefont {Fu}\ and\ \citenamefont {Kane}(2009)}]{Fu2009}%
  \BibitemOpen
  \bibfield  {author} {\bibinfo {author} {\bibfnamefont {L.}~\bibnamefont
  {Fu}}\ and\ \bibinfo {author} {\bibfnamefont {C.~L.}\ \bibnamefont {Kane}},\
  }\href {\doibase 10.1103/PhysRevB.79.161408} {\bibfield  {journal} {\bibinfo
  {journal} {Phys. Rev. B}\ }\textbf {\bibinfo {volume} {79}},\ \bibinfo
  {pages} {161408(R)} (\bibinfo {year} {2009})}\BibitemShut {NoStop}%
\bibitem [{\citenamefont {Mourik}\ \emph {et~al.}(2012)\citenamefont {Mourik},
  \citenamefont {Zuo}, \citenamefont {Frolov}, \citenamefont {Plissard},
  \citenamefont {Bakkers},\ and\ \citenamefont {Kouwenhoven}}]{Mourik2012}%
  \BibitemOpen
  \bibfield  {author} {\bibinfo {author} {\bibfnamefont {V.}~\bibnamefont
  {Mourik}}, \bibinfo {author} {\bibfnamefont {K.}~\bibnamefont {Zuo}},
  \bibinfo {author} {\bibfnamefont {S.~M.}\ \bibnamefont {Frolov}}, \bibinfo
  {author} {\bibfnamefont {S.~R.}\ \bibnamefont {Plissard}}, \bibinfo {author}
  {\bibfnamefont {E.~P. A.~M.}\ \bibnamefont {Bakkers}}, \ and\ \bibinfo
  {author} {\bibfnamefont {L.~P.}\ \bibnamefont {Kouwenhoven}},\ }\href
  {\doibase 10.1126/science.1222360} {\bibfield  {journal} {\bibinfo  {journal}
  {Science}\ }\textbf {\bibinfo {volume} {336}},\ \bibinfo {pages} {1003}
  (\bibinfo {year} {2012})}\BibitemShut {NoStop}%
\bibitem [{\citenamefont {Deng}\ \emph {et~al.}(2012)\citenamefont {Deng},
  \citenamefont {Yu}, \citenamefont {Huang}, \citenamefont {Larsson},
  \citenamefont {Caroff},\ and\ \citenamefont {Xu}}]{Deng2012}%
  \BibitemOpen
  \bibfield  {author} {\bibinfo {author} {\bibfnamefont {M.~T.}\ \bibnamefont
  {Deng}}, \bibinfo {author} {\bibfnamefont {C.~L.}\ \bibnamefont {Yu}},
  \bibinfo {author} {\bibfnamefont {G.~Y.}\ \bibnamefont {Huang}}, \bibinfo
  {author} {\bibfnamefont {M.}~\bibnamefont {Larsson}}, \bibinfo {author}
  {\bibfnamefont {P.}~\bibnamefont {Caroff}}, \ and\ \bibinfo {author}
  {\bibfnamefont {H.~Q.}\ \bibnamefont {Xu}},\ }\href {\doibase
  10.1021/nl303758w} {\bibfield  {journal} {\bibinfo  {journal} {Nano Lett.}\
  }\textbf {\bibinfo {volume} {12}},\ \bibinfo {pages} {6414} (\bibinfo {year}
  {2012})}\BibitemShut {NoStop}%
\bibitem [{\citenamefont {Das}\ \emph {et~al.}(2012)\citenamefont {Das},
  \citenamefont {Ronen}, \citenamefont {Most}, \citenamefont {Oreg},
  \citenamefont {Heiblum},\ and\ \citenamefont {Shtrikman}}]{Das2012}%
  \BibitemOpen
  \bibfield  {author} {\bibinfo {author} {\bibfnamefont {A.}~\bibnamefont
  {Das}}, \bibinfo {author} {\bibfnamefont {Y.}~\bibnamefont {Ronen}}, \bibinfo
  {author} {\bibfnamefont {Y.}~\bibnamefont {Most}}, \bibinfo {author}
  {\bibfnamefont {Y.}~\bibnamefont {Oreg}}, \bibinfo {author} {\bibfnamefont
  {M.}~\bibnamefont {Heiblum}}, \ and\ \bibinfo {author} {\bibfnamefont
  {H.}~\bibnamefont {Shtrikman}},\ }\href
  {http://www.nature.com/nphys/journal/v8/n12/full/nphys2479.html} {\bibfield
  {journal} {\bibinfo  {journal} {Nat. Phys.}\ }\textbf {\bibinfo {volume}
  {8}},\ \bibinfo {pages} {887} (\bibinfo {year} {2012})}\BibitemShut {NoStop}%
\bibitem [{\citenamefont {Lee}\ \emph {et~al.}(2012)\citenamefont {Lee},
  \citenamefont {Jiang}, \citenamefont {Aguado}, \citenamefont {Katsaros},
  \citenamefont {Lieber},\ and\ \citenamefont {De~Franceschi}}]{Lee2012}%
  \BibitemOpen
  \bibfield  {author} {\bibinfo {author} {\bibfnamefont {E.~J.~H.}\
  \bibnamefont {Lee}}, \bibinfo {author} {\bibfnamefont {X.}~\bibnamefont
  {Jiang}}, \bibinfo {author} {\bibfnamefont {R.}~\bibnamefont {Aguado}},
  \bibinfo {author} {\bibfnamefont {G.}~\bibnamefont {Katsaros}}, \bibinfo
  {author} {\bibfnamefont {C.~M.}\ \bibnamefont {Lieber}}, \ and\ \bibinfo
  {author} {\bibfnamefont {S.}~\bibnamefont {De~Franceschi}},\ }\href {\doibase
  10.1103/PhysRevLett.109.186802} {\bibfield  {journal} {\bibinfo  {journal}
  {Phys. Rev. Lett.}\ }\textbf {\bibinfo {volume} {109}},\ \bibinfo {pages}
  {186802} (\bibinfo {year} {2012})}\BibitemShut {NoStop}%
\bibitem [{\citenamefont {Finck}\ \emph {et~al.}(2013)\citenamefont {Finck},
  \citenamefont {Van~Harlingen}, \citenamefont {Mohseni}, \citenamefont
  {Jung},\ and\ \citenamefont {Li}}]{Finck2013}%
  \BibitemOpen
  \bibfield  {author} {\bibinfo {author} {\bibfnamefont {A.~D.~K.}\
  \bibnamefont {Finck}}, \bibinfo {author} {\bibfnamefont {D.~J.}\ \bibnamefont
  {Van~Harlingen}}, \bibinfo {author} {\bibfnamefont {P.~K.}\ \bibnamefont
  {Mohseni}}, \bibinfo {author} {\bibfnamefont {K.}~\bibnamefont {Jung}}, \
  and\ \bibinfo {author} {\bibfnamefont {X.}~\bibnamefont {Li}},\ }\href
  {\doibase 10.1103/PhysRevLett.110.126406} {\bibfield  {journal} {\bibinfo
  {journal} {Phys. Rev. Lett.}\ }\textbf {\bibinfo {volume} {110}},\ \bibinfo
  {pages} {126406} (\bibinfo {year} {2013})}\BibitemShut {NoStop}%
\bibitem [{\citenamefont {Churchill}\ \emph {et~al.}(2013)\citenamefont
  {Churchill}, \citenamefont {Fatemi}, \citenamefont {Grove-Rasmussen},
  \citenamefont {Deng}, \citenamefont {Caroff}, \citenamefont {Xu},\ and\
  \citenamefont {Marcus}}]{Churchill2013}%
  \BibitemOpen
  \bibfield  {author} {\bibinfo {author} {\bibfnamefont {H.~O.~H.}\
  \bibnamefont {Churchill}}, \bibinfo {author} {\bibfnamefont {V.}~\bibnamefont
  {Fatemi}}, \bibinfo {author} {\bibfnamefont {K.}~\bibnamefont
  {Grove-Rasmussen}}, \bibinfo {author} {\bibfnamefont {M.~T.}\ \bibnamefont
  {Deng}}, \bibinfo {author} {\bibfnamefont {P.}~\bibnamefont {Caroff}},
  \bibinfo {author} {\bibfnamefont {H.~Q.}\ \bibnamefont {Xu}}, \ and\ \bibinfo
  {author} {\bibfnamefont {C.~M.}\ \bibnamefont {Marcus}},\ }\href {\doibase
  10.1103/PhysRevB.87.241401} {\bibfield  {journal} {\bibinfo  {journal} {Phys.
  Rev. B}\ }\textbf {\bibinfo {volume} {87}},\ \bibinfo {pages} {241401(R)}
  (\bibinfo {year} {2013})}\BibitemShut {NoStop}%
\bibitem [{\citenamefont {Goldhaber-Gordon}\ \emph {et~al.}(1998)\citenamefont
  {Goldhaber-Gordon}, \citenamefont {Shtrikman}, \citenamefont {Mahalu},
  \citenamefont {Abusch-Magder}, \citenamefont {Meirav},\ and\ \citenamefont
  {Kastner}}]{Goldhaber1998}%
  \BibitemOpen
  \bibfield  {author} {\bibinfo {author} {\bibfnamefont {D.}~\bibnamefont
  {Goldhaber-Gordon}}, \bibinfo {author} {\bibfnamefont {H.}~\bibnamefont
  {Shtrikman}}, \bibinfo {author} {\bibfnamefont {D.}~\bibnamefont {Mahalu}},
  \bibinfo {author} {\bibfnamefont {D.}~\bibnamefont {Abusch-Magder}}, \bibinfo
  {author} {\bibfnamefont {U.}~\bibnamefont {Meirav}}, \ and\ \bibinfo {author}
  {\bibfnamefont {M.}~\bibnamefont {Kastner}},\ }\href
  {https://www.nature.com/nature/journal/v391/n6663/full/391156a0.html}
  {\bibfield  {journal} {\bibinfo  {journal} {Nature}\ }\textbf {\bibinfo
  {volume} {391}},\ \bibinfo {pages} {157} (\bibinfo {year}
  {1998})}\BibitemShut {NoStop}%
\bibitem [{\citenamefont {Cronenwett}\ \emph {et~al.}(1998)\citenamefont
  {Cronenwett}, \citenamefont {Oosterkamp},\ and\ \citenamefont
  {Kouwenhoven}}]{Cronenwett1998}%
  \BibitemOpen
  \bibfield  {author} {\bibinfo {author} {\bibfnamefont {S.~M.}\ \bibnamefont
  {Cronenwett}}, \bibinfo {author} {\bibfnamefont {T.~H.}\ \bibnamefont
  {Oosterkamp}}, \ and\ \bibinfo {author} {\bibfnamefont {L.~P.}\ \bibnamefont
  {Kouwenhoven}},\ }\href {http://science.sciencemag.org/content/281/5376/540}
  {\bibfield  {journal} {\bibinfo  {journal} {Science}\ }\textbf {\bibinfo
  {volume} {281}},\ \bibinfo {pages} {540} (\bibinfo {year}
  {1998})}\BibitemShut {NoStop}%
\bibitem [{\citenamefont {Golubov}\ \emph {et~al.}(2009)\citenamefont
  {Golubov}, \citenamefont {Brinkman}, \citenamefont {Tanaka}, \citenamefont
  {Mazin},\ and\ \citenamefont {Dolgov}}]{Golubov2009}%
  \BibitemOpen
  \bibfield  {author} {\bibinfo {author} {\bibfnamefont {A.~A.}\ \bibnamefont
  {Golubov}}, \bibinfo {author} {\bibfnamefont {A.}~\bibnamefont {Brinkman}},
  \bibinfo {author} {\bibfnamefont {Y.}~\bibnamefont {Tanaka}}, \bibinfo
  {author} {\bibfnamefont {I.~I.}\ \bibnamefont {Mazin}}, \ and\ \bibinfo
  {author} {\bibfnamefont {O.~V.}\ \bibnamefont {Dolgov}},\ }\href {\doibase
  10.1103/PhysRevLett.103.077003} {\bibfield  {journal} {\bibinfo  {journal}
  {Phys. Rev. Lett.}\ }\textbf {\bibinfo {volume} {103}},\ \bibinfo {pages}
  {077003} (\bibinfo {year} {2009})}\BibitemShut {NoStop}%
\bibitem [{\citenamefont {Holleitner}\ \emph {et~al.}(2001)\citenamefont
  {Holleitner}, \citenamefont {Decker}, \citenamefont {Qin}, \citenamefont
  {Eberl},\ and\ \citenamefont {Blick}}]{Holleitner2001}%
  \BibitemOpen
  \bibfield  {author} {\bibinfo {author} {\bibfnamefont {A.~W.}\ \bibnamefont
  {Holleitner}}, \bibinfo {author} {\bibfnamefont {C.~R.}\ \bibnamefont
  {Decker}}, \bibinfo {author} {\bibfnamefont {H.}~\bibnamefont {Qin}},
  \bibinfo {author} {\bibfnamefont {K.}~\bibnamefont {Eberl}}, \ and\ \bibinfo
  {author} {\bibfnamefont {R.~H.}\ \bibnamefont {Blick}},\ }\href {\doibase
  10.1103/PhysRevLett.87.256802} {\bibfield  {journal} {\bibinfo  {journal}
  {Phys. Rev. Lett.}\ }\textbf {\bibinfo {volume} {87}},\ \bibinfo {pages}
  {256802} (\bibinfo {year} {2001})}\BibitemShut {NoStop}%
\bibitem [{\citenamefont {Holleitner}\ \emph {et~al.}(2002)\citenamefont
  {Holleitner}, \citenamefont {Blick}, \citenamefont {H{\"u}ttel},
  \citenamefont {Eberl},\ and\ \citenamefont {Kotthaus}}]{Holleitner2002}%
  \BibitemOpen
  \bibfield  {author} {\bibinfo {author} {\bibfnamefont {A.~W.}\ \bibnamefont
  {Holleitner}}, \bibinfo {author} {\bibfnamefont {R.~H.}\ \bibnamefont
  {Blick}}, \bibinfo {author} {\bibfnamefont {A.~K.}\ \bibnamefont
  {H{\"u}ttel}}, \bibinfo {author} {\bibfnamefont {K.}~\bibnamefont {Eberl}}, \
  and\ \bibinfo {author} {\bibfnamefont {J.~P.}\ \bibnamefont {Kotthaus}},\
  }\href {http://science.sciencemag.org/content/297/5578/70} {\bibfield
  {journal} {\bibinfo  {journal} {Science}\ }\textbf {\bibinfo {volume}
  {297}},\ \bibinfo {pages} {70} (\bibinfo {year} {2002})}\BibitemShut
  {NoStop}%
\bibitem [{\citenamefont {Shangguan}\ \emph {et~al.}(2001)\citenamefont
  {Shangguan}, \citenamefont {Au~Yeung}, \citenamefont {Yu},\ and\
  \citenamefont {Kam}}]{Shangguan2001}%
  \BibitemOpen
  \bibfield  {author} {\bibinfo {author} {\bibfnamefont {W.~Z.}\ \bibnamefont
  {Shangguan}}, \bibinfo {author} {\bibfnamefont {T.~C.}\ \bibnamefont
  {Au~Yeung}}, \bibinfo {author} {\bibfnamefont {Y.~B.}\ \bibnamefont {Yu}}, \
  and\ \bibinfo {author} {\bibfnamefont {C.~H.}\ \bibnamefont {Kam}},\ }\href
  {\doibase 10.1103/PhysRevB.63.235323} {\bibfield  {journal} {\bibinfo
  {journal} {Phys. Rev. B}\ }\textbf {\bibinfo {volume} {63}},\ \bibinfo
  {pages} {235323} (\bibinfo {year} {2001})}\BibitemShut {NoStop}%
\bibitem [{\citenamefont {Orellana}\ \emph {et~al.}(2003)\citenamefont
  {Orellana}, \citenamefont {Dom\'{\i}nguez-Adame}, \citenamefont {G\'omez},\
  and\ \citenamefont {Ladr\'on~de Guevara}}]{Orellana2003}%
  \BibitemOpen
  \bibfield  {author} {\bibinfo {author} {\bibfnamefont {P.~A.}\ \bibnamefont
  {Orellana}}, \bibinfo {author} {\bibfnamefont {F.}~\bibnamefont
  {Dom\'{\i}nguez-Adame}}, \bibinfo {author} {\bibfnamefont {I.}~\bibnamefont
  {G\'omez}}, \ and\ \bibinfo {author} {\bibfnamefont {M.~L.}\ \bibnamefont
  {Ladr\'on~de Guevara}},\ }\href {\doibase 10.1103/PhysRevB.67.085321}
  {\bibfield  {journal} {\bibinfo  {journal} {Phys. Rev. B}\ }\textbf {\bibinfo
  {volume} {67}},\ \bibinfo {pages} {085321} (\bibinfo {year}
  {2003})}\BibitemShut {NoStop}%
\bibitem [{\citenamefont {Fano}(1961)}]{Fano1961}%
  \BibitemOpen
  \bibfield  {author} {\bibinfo {author} {\bibfnamefont {U.}~\bibnamefont
  {Fano}},\ }\href {\doibase 10.1103/PhysRev.124.1866} {\bibfield  {journal}
  {\bibinfo  {journal} {Phys. Rev.}\ }\textbf {\bibinfo {volume} {124}},\
  \bibinfo {pages} {1866} (\bibinfo {year} {1961})}\BibitemShut {NoStop}%
\bibitem [{\citenamefont {Miroshnichenko}\ \emph {et~al.}(2010)\citenamefont
  {Miroshnichenko}, \citenamefont {Flach},\ and\ \citenamefont
  {Kivshar}}]{Miroshnichenko2010}%
  \BibitemOpen
  \bibfield  {author} {\bibinfo {author} {\bibfnamefont {A.~E.}\ \bibnamefont
  {Miroshnichenko}}, \bibinfo {author} {\bibfnamefont {S.}~\bibnamefont
  {Flach}}, \ and\ \bibinfo {author} {\bibfnamefont {Y.~S.}\ \bibnamefont
  {Kivshar}},\ }\href {\doibase 10.1103/RevModPhys.82.2257} {\bibfield
  {journal} {\bibinfo  {journal} {Rev. Mod. Phys.}\ }\textbf {\bibinfo {volume}
  {82}},\ \bibinfo {pages} {2257} (\bibinfo {year} {2010})}\BibitemShut
  {NoStop}%
\bibitem [{\citenamefont {Vernek}\ \emph {et~al.}(2014)\citenamefont {Vernek},
  \citenamefont {Penteado}, \citenamefont {Seridonio},\ and\ \citenamefont
  {Egues}}]{Vernek2014}%
  \BibitemOpen
  \bibfield  {author} {\bibinfo {author} {\bibfnamefont {E.}~\bibnamefont
  {Vernek}}, \bibinfo {author} {\bibfnamefont {P.~H.}\ \bibnamefont
  {Penteado}}, \bibinfo {author} {\bibfnamefont {A.~C.}\ \bibnamefont
  {Seridonio}}, \ and\ \bibinfo {author} {\bibfnamefont {J.~C.}\ \bibnamefont
  {Egues}},\ }\href {\doibase 10.1103/PhysRevB.89.165314} {\bibfield  {journal}
  {\bibinfo  {journal} {Phys. Rev. B}\ }\textbf {\bibinfo {volume} {89}},\
  \bibinfo {pages} {165314} (\bibinfo {year} {2014})}\BibitemShut {NoStop}%
\bibitem [{\citenamefont {Ruiz-Tijerina}\ \emph {et~al.}(2015)\citenamefont
  {Ruiz-Tijerina}, \citenamefont {Vernek}, \citenamefont {Dias~da Silva},\ and\
  \citenamefont {Egues}}]{Ruiz2015}%
  \BibitemOpen
  \bibfield  {author} {\bibinfo {author} {\bibfnamefont {D.~A.}\ \bibnamefont
  {Ruiz-Tijerina}}, \bibinfo {author} {\bibfnamefont {E.}~\bibnamefont
  {Vernek}}, \bibinfo {author} {\bibfnamefont {L.~G. G.~V.}\ \bibnamefont
  {Dias~da Silva}}, \ and\ \bibinfo {author} {\bibfnamefont {J.~C.}\
  \bibnamefont {Egues}},\ }\href {\doibase 10.1103/PhysRevB.91.115435}
  {\bibfield  {journal} {\bibinfo  {journal} {Phys. Rev. B}\ }\textbf {\bibinfo
  {volume} {91}},\ \bibinfo {pages} {115435} (\bibinfo {year}
  {2015})}\BibitemShut {NoStop}%
\bibitem [{\citenamefont {Gong}\ \emph {et~al.}(2014)\citenamefont {Gong},
  \citenamefont {Zhang}, \citenamefont {Li}, \citenamefont {Yi},\ and\
  \citenamefont {Zheng}}]{Gong2014}%
  \BibitemOpen
  \bibfield  {author} {\bibinfo {author} {\bibfnamefont {W.-J.}\ \bibnamefont
  {Gong}}, \bibinfo {author} {\bibfnamefont {S.-F.}\ \bibnamefont {Zhang}},
  \bibinfo {author} {\bibfnamefont {Z.-C.}\ \bibnamefont {Li}}, \bibinfo
  {author} {\bibfnamefont {G.}~\bibnamefont {Yi}}, \ and\ \bibinfo {author}
  {\bibfnamefont {Y.-S.}\ \bibnamefont {Zheng}},\ }\href {\doibase
  10.1103/PhysRevB.89.245413} {\bibfield  {journal} {\bibinfo  {journal} {Phys.
  Rev. B}\ }\textbf {\bibinfo {volume} {89}},\ \bibinfo {pages} {245413}
  (\bibinfo {year} {2014})}\BibitemShut {NoStop}%
\bibitem [{\citenamefont {Meir}\ and\ \citenamefont
  {Wingreen}(1992)}]{Meir1992}%
  \BibitemOpen
  \bibfield  {author} {\bibinfo {author} {\bibfnamefont {Y.}~\bibnamefont
  {Meir}}\ and\ \bibinfo {author} {\bibfnamefont {N.~S.}\ \bibnamefont
  {Wingreen}},\ }\href {\doibase 10.1103/PhysRevLett.68.2512} {\bibfield
  {journal} {\bibinfo  {journal} {Phys. Rev. Lett.}\ }\textbf {\bibinfo
  {volume} {68}},\ \bibinfo {pages} {2512} (\bibinfo {year}
  {1992})}\BibitemShut {NoStop}%
\bibitem [{\citenamefont {Liu}\ and\ \citenamefont {Baranger}(2011)}]{Liu2011}%
  \BibitemOpen
  \bibfield  {author} {\bibinfo {author} {\bibfnamefont {D.~E.}\ \bibnamefont
  {Liu}}\ and\ \bibinfo {author} {\bibfnamefont {H.~U.}\ \bibnamefont
  {Baranger}},\ }\href {\doibase 10.1103/PhysRevB.84.201308} {\bibfield
  {journal} {\bibinfo  {journal} {Phys. Rev. B}\ }\textbf {\bibinfo {volume}
  {84}},\ \bibinfo {pages} {201308(R)} (\bibinfo {year} {2011})}\BibitemShut
  {NoStop}%
\bibitem [{\citenamefont {Deng}\ \emph {et~al.}(2016)\citenamefont {Deng},
  \citenamefont {Vaitiekenas}, \citenamefont {Hansen}, \citenamefont {Danon},
  \citenamefont {Leijnse}, \citenamefont {Flensberg}, \citenamefont {Nyg{\r
  a}rd}, \citenamefont {Krogstrup},\ and\ \citenamefont {Marcus}}]{Deng2016}%
  \BibitemOpen
  \bibfield  {author} {\bibinfo {author} {\bibfnamefont {M.~T.}\ \bibnamefont
  {Deng}}, \bibinfo {author} {\bibfnamefont {S.}~\bibnamefont {Vaitiekenas}},
  \bibinfo {author} {\bibfnamefont {E.~B.}\ \bibnamefont {Hansen}}, \bibinfo
  {author} {\bibfnamefont {J.}~\bibnamefont {Danon}}, \bibinfo {author}
  {\bibfnamefont {M.}~\bibnamefont {Leijnse}}, \bibinfo {author} {\bibfnamefont
  {K.}~\bibnamefont {Flensberg}}, \bibinfo {author} {\bibfnamefont
  {J.}~\bibnamefont {Nyg{\r a}rd}}, \bibinfo {author} {\bibfnamefont
  {P.}~\bibnamefont {Krogstrup}}, \ and\ \bibinfo {author} {\bibfnamefont
  {C.~M.}\ \bibnamefont {Marcus}},\ }\href {\doibase 10.1126/science.aaf3961}
  {\bibfield  {journal} {\bibinfo  {journal} {Science}\ }\textbf {\bibinfo
  {volume} {354}},\ \bibinfo {pages} {1557} (\bibinfo {year}
  {2016})}\BibitemShut {NoStop}%
\bibitem [{\citenamefont {Hsu}\ \emph {et~al.}(2016{\natexlab{a}})\citenamefont
  {Hsu}, \citenamefont {Zhen}, \citenamefont {Stone}, \citenamefont
  {Joannopoulos},\ and\ \citenamefont {Solja{\v{c}}i{\'c}}}]{bics}%
  \BibitemOpen
  \bibfield  {author} {\bibinfo {author} {\bibfnamefont {C.~W.}\ \bibnamefont
  {Hsu}}, \bibinfo {author} {\bibfnamefont {B.}~\bibnamefont {Zhen}}, \bibinfo
  {author} {\bibfnamefont {A.~D.}\ \bibnamefont {Stone}}, \bibinfo {author}
  {\bibfnamefont {J.~D.}\ \bibnamefont {Joannopoulos}}, \ and\ \bibinfo
  {author} {\bibfnamefont {M.}~\bibnamefont {Solja{\v{c}}i{\'c}}},\ }\href
  {https://www.nature.com/articles/natrevmats201648?WT.feed_name=subjects_quantum-mechanics}
  {\bibfield  {journal} {\bibinfo  {journal} {Nature Reviews Materials}\
  }\textbf {\bibinfo {volume} {1}},\ \bibinfo {pages} {16048} (\bibinfo {year}
  {2016}{\natexlab{a}})}\BibitemShut {NoStop}%
\bibitem [{\citenamefont {{von Neumann}}\ and\ \citenamefont
  {{Wigner}}(1929)}]{vonNeumann1929}%
  \BibitemOpen
  \bibfield  {author} {\bibinfo {author} {\bibfnamefont {J.}~\bibnamefont {{von
  Neumann}}}\ and\ \bibinfo {author} {\bibfnamefont {E.}~\bibnamefont
  {{Wigner}}},\ }\href@noop {} {\bibfield  {journal} {\bibinfo  {journal}
  {Phys. Z}\ }\textbf {\bibinfo {volume} {30}},\ \bibinfo {pages} {467}
  (\bibinfo {year} {1929})}\BibitemShut {NoStop}%
\bibitem [{\citenamefont {Hsu}\ \emph {et~al.}(2016{\natexlab{b}})\citenamefont
  {Hsu}, \citenamefont {Zhen}, \citenamefont {Stone}, \citenamefont
  {Joannopoulos},\ and\ \citenamefont {Solja{\v{c}}i{\'c}}}]{Hsu2016}%
  \BibitemOpen
  \bibfield  {author} {\bibinfo {author} {\bibfnamefont {C.~W.}\ \bibnamefont
  {Hsu}}, \bibinfo {author} {\bibfnamefont {B.}~\bibnamefont {Zhen}}, \bibinfo
  {author} {\bibfnamefont {A.~D.}\ \bibnamefont {Stone}}, \bibinfo {author}
  {\bibfnamefont {J.~D.}\ \bibnamefont {Joannopoulos}}, \ and\ \bibinfo
  {author} {\bibfnamefont {M.}~\bibnamefont {Solja{\v{c}}i{\'c}}},\ }\href
  {https://www.nature.com/articles/natrevmats201648} {\bibfield  {journal}
  {\bibinfo  {journal} {Nat. Rev. Mat.}\ }\textbf {\bibinfo {volume} {1}},\
  \bibinfo {pages} {16048} (\bibinfo {year} {2016}{\natexlab{b}})}\BibitemShut
  {NoStop}%
\bibitem [{\citenamefont {Ramos}\ and\ \citenamefont
  {Orellana}(2014)}]{Ramos2014}%
  \BibitemOpen
  \bibfield  {author} {\bibinfo {author} {\bibfnamefont {J.~P.}\ \bibnamefont
  {Ramos}}\ and\ \bibinfo {author} {\bibfnamefont {P.~A.}\ \bibnamefont
  {Orellana}},\ }\href
  {http://www.sciencedirect.com/science/article/pii/S0921452614005900}
  {\bibfield  {journal} {\bibinfo  {journal} {Phys. B}\ }\textbf {\bibinfo
  {volume} {455}},\ \bibinfo {pages} {66} (\bibinfo {year} {2014})}\BibitemShut
  {NoStop}%
\bibitem [{\citenamefont {Guessi}\ \emph {et~al.}(2017)\citenamefont {Guessi},
  \citenamefont {Dessotti}, \citenamefont {Marques}, \citenamefont {Ricco},
  \citenamefont {Pereira}, \citenamefont {Menegasso}, \citenamefont
  {de~Souza},\ and\ \citenamefont {Seridonio}}]{Guessi2017}%
  \BibitemOpen
  \bibfield  {author} {\bibinfo {author} {\bibfnamefont {L.~H.}\ \bibnamefont
  {Guessi}}, \bibinfo {author} {\bibfnamefont {F.~A.}\ \bibnamefont
  {Dessotti}}, \bibinfo {author} {\bibfnamefont {Y.}~\bibnamefont {Marques}},
  \bibinfo {author} {\bibfnamefont {L.~S.}\ \bibnamefont {Ricco}}, \bibinfo
  {author} {\bibfnamefont {G.~M.}\ \bibnamefont {Pereira}}, \bibinfo {author}
  {\bibfnamefont {P.}~\bibnamefont {Menegasso}}, \bibinfo {author}
  {\bibfnamefont {M.}~\bibnamefont {de~Souza}}, \ and\ \bibinfo {author}
  {\bibfnamefont {A.~C.}\ \bibnamefont {Seridonio}},\ }\href {\doibase
  10.1103/PhysRevB.96.041114} {\bibfield  {journal} {\bibinfo  {journal} {Phys.
  Rev. B}\ }\textbf {\bibinfo {volume} {96}},\ \bibinfo {pages} {041114(R)}
  (\bibinfo {year} {2017})}\BibitemShut {NoStop}%
\bibitem [{\citenamefont {Ricco}\ \emph {et~al.}(2016)\citenamefont {Ricco},
  \citenamefont {Marques}, \citenamefont {Dessotti}, \citenamefont {Machado},
  \citenamefont {de~Souza},\ and\ \citenamefont {Seridonio}}]{Ricco2016}%
  \BibitemOpen
  \bibfield  {author} {\bibinfo {author} {\bibfnamefont {L.~S.}\ \bibnamefont
  {Ricco}}, \bibinfo {author} {\bibfnamefont {Y.}~\bibnamefont {Marques}},
  \bibinfo {author} {\bibfnamefont {F.~A.}\ \bibnamefont {Dessotti}}, \bibinfo
  {author} {\bibfnamefont {R.~S.}\ \bibnamefont {Machado}}, \bibinfo {author}
  {\bibfnamefont {M.}~\bibnamefont {de~Souza}}, \ and\ \bibinfo {author}
  {\bibfnamefont {A.~C.}\ \bibnamefont {Seridonio}},\ }\href {\doibase
  10.1103/PhysRevB.93.165116} {\bibfield  {journal} {\bibinfo  {journal} {Phys.
  Rev. B}\ }\textbf {\bibinfo {volume} {93}},\ \bibinfo {pages} {165116}
  (\bibinfo {year} {2016})}\BibitemShut {NoStop}%
\bibitem [{\citenamefont {Zambrano}\ \emph {et~al.}(2018)\citenamefont
  {Zambrano}, \citenamefont {Ramos-Andrade},\ and\ \citenamefont
  {Orellana}}]{Zambrano2018}%
  \BibitemOpen
  \bibfield  {author} {\bibinfo {author} {\bibfnamefont {D.}~\bibnamefont
  {Zambrano}}, \bibinfo {author} {\bibfnamefont {J.~P.}\ \bibnamefont
  {Ramos-Andrade}}, \ and\ \bibinfo {author} {\bibfnamefont {P.~A.}\
  \bibnamefont {Orellana}},\ }\href {\doibase 10.1088/1361-648X/aad7ca}
  {\bibfield  {journal} {\bibinfo  {journal} {J. Phys.: Condens. Matter}\
  }\textbf {\bibinfo {volume} {30}} (\bibinfo {year} {2018}),\
  10.1088/1361-648X/aad7ca}\BibitemShut {NoStop}%
\bibitem [{\citenamefont {Alicea}\ \emph {et~al.}(2011)\citenamefont {Alicea},
  \citenamefont {Oreg}, \citenamefont {Refael}, \citenamefont {Von~Oppen},\
  and\ \citenamefont {Fisher}}]{Alicea2011}%
  \BibitemOpen
  \bibfield  {author} {\bibinfo {author} {\bibfnamefont {J.}~\bibnamefont
  {Alicea}}, \bibinfo {author} {\bibfnamefont {Y.}~\bibnamefont {Oreg}},
  \bibinfo {author} {\bibfnamefont {G.}~\bibnamefont {Refael}}, \bibinfo
  {author} {\bibfnamefont {F.}~\bibnamefont {Von~Oppen}}, \ and\ \bibinfo
  {author} {\bibfnamefont {M.~P.}\ \bibnamefont {Fisher}},\ }\href
  {http://www.nature.com/nphys/journal/v7/n5/abs/nphys1915.html} {\bibfield
  {journal} {\bibinfo  {journal} {Nature Physics}\ }\textbf {\bibinfo {volume}
  {7}},\ \bibinfo {pages} {412} (\bibinfo {year} {2011})}\BibitemShut {NoStop}%
\end{thebibliography}%

\end{document}